\documentclass[namedreferences]{SolarPhysics}
\usepackage[optionalrh]{spr-sola-addons} 
\usepackage{graphicx}
\usepackage{url}             

\begin{document}
\begin{article}
\begin{opening}

\title{An Extreme Solar Event of 20 January 2005: Properties of the Flare
and the Origin of Energetic Particles}

\author{V.V.~\surname{Grechnev}$^{1}$\sep
        V.G.~\surname{Kurt}$^{2}$\sep
        I.M.~\surname{Chertok}$^{3}$\sep
        A.M.~\surname{Uralov}$^{1}$\sep
        H.~\surname{Nakajima}$^{4}$\sep
        A.T.~\surname{Altyntsev}$^{1}$\sep
        A.V.~\surname{Belov}$^{3}$\sep
        B.Yu.~\surname{Yushkov}$^{2}$\sep
        S.N.~\surname{Kuznetsov}$^{2}$\thanks{Deceased 17 May 2007}\sep
        L.K.~\surname{Kashapova}$^{1}$\sep
        N.S.~\surname{Meshalkina}$^{1}$\sep
        N.P.~\surname{Prestage}$^{5}$}

 \runningauthor{Grechnev \textit{et al.}}
 \runningtitle{An Extreme Solar Event of 20 January 2005}

\institute{$^{1}$ Institute of Solar-Terrestrial Physics SB RAS,
Lermontov St.\ 126A, Irkutsk 664033, Russia
                  \email{grechnev@iszf.irk.ru}\\
           $^{2}$ Skobeltsyn Institute of Nuclear Physics Moscow Lomonosov State
                  University, Moscow, 119992 Russia\\
           $^{3}$ Pushkov Institute of Terrestrial Magnetism,
                  Ionosphere and Radiowave Propagation (IZMIRAN), Troitsk, Moscow
                  Region, 142190 Russia\\
           $^{4}$ Nobeyama Radio Observatory, Minamimaki, Minamisaku,
                  Nagano 384-1305, Japan\\
           $^{5}$ IPS Radio and Space Services Culgoora Solar Observatory, Australia\\
             }

\date{Received ; accepted }

\begin{abstract}
The famous extreme solar and particle event of 20 January 2005 is
analyzed from two perspectives. Firstly, using multi-spectral
data, we study temporal, spectral, and spatial features of the
main phase of the flare, when the strongest emissions from
microwaves up to 200 MeV gamma-rays were observed. Secondly, we
relate our results to a long-standing controversy on the origin of
solar energetic particles (SEP) arriving at Earth, \textit{i.e.},
acceleration in flares, or shocks ahead of coronal mass ejections
(CMEs). Our analysis shows that all electromagnetic emissions from
microwaves up to 2.22 MeV line gamma-rays during the main flare
phase originated within a compact structure located just above
sunspot umbrae. In particular, a huge ($\approx 10^5$ sfu) radio
burst with a high frequency maximum at 30 GHz was observed,
indicating the presence of a large number of energetic electrons
in very strong magnetic fields. Thus, protons and electrons
responsible for various flare emissions during its main phase were
accelerated within the magnetic field of the active region. The
leading, impulsive parts of the ground-level enhancement (GLE),
and highest-energy gamma-rays identified with $\pi^0$-decay
emission, are similar and closely correspond in time. The origin
of the $\pi^0$-decay gamma-rays is argued to be the same as that
of lower-energy emissions, although this is not proven. On the
other hand, we estimate the sky-plane speed of the CME to be
2\,000\,--\,2\,600 km\thinspace s$^{-1}$, \textit{i.e.}, high, but
of the same order as preceding non-GLE-related CMEs from the same
active region. Hence, the flare itself rather than the CME appears
to determine the extreme nature of this event. We therefore
conclude that the acceleration, at least, to sub-relativistic
energies, of electrons and protons, responsible for both the major
flare emissions and the leading spike of SEP/GLE by 07~UT, are
likely to have occurred nearly simultaneously within the flare
region. However, our analysis does not rule out a probable
contribution from particles accelerated in the CME-driven shock
for the leading GLE spike, which seemed to dominate at later
stages of the SEP event.
\end{abstract}


\end{opening}

\section{Introduction}

The solar eruptive-flare event of 20 January 2005 has attracted
great attention from the solar and solar-terrestrial community due to
its outstanding characteristics (see, \textit{e.g.},
http:/\negthinspace/creme96.nrl.navy.mil/20Jan05/). Despite its
occurrence at the deep descending phase of the solar cycle, the
event was characterized, in particular, by strong gamma-ray emission
with a photon energy up to at least 200 MeV; by one of the
strongest microwave bursts with a spectral maximum frequency
$\approx 30$~GHz, by a fast halo coronal mass ejection (CME), and
was accompanied by a very hard-spectrum solar energetic particle
(SEP) flux near Earth including the second largest ground-level
enhancement (GLE) of cosmic ray intensity in observational
history.

Many studies are devoted to this extreme event, but they mainly
concern the cosmic ray aspect. The central points of these studies
are the features and origin of the SEP/GLE. Additionally, a
long-standing discussion over the origin of accelerated protons in
solar events, \textit{i.e.}, CME-driven shock-acceleration or
flare-acceleration, has been strengthened. In particular,
Gopalswamy \textit{et al.} (2005) estimated that the CME had the
largest sky-plane speed exceeding 3000 km\thinspace s$^{-1}$ and
concluded that the GLE had a shock origin. On the other hand,
Simnett (2006, 2007) considered the relative timing of various
manifestations in this event and several other factors, and stated
that the CME was not responsible for the relativistic ion
acceleration. Kuznetsov \textit{et al.} (2005a,b, 2006a,b, 2007)
also argued in favor of flare-related proton acceleration in this
and other events.

These contradicting opinions reflect a long-standing controversy
over different viewpoints of the origin of high-energy protons
related to solar events [see Cliver (2000) for a historical
review]. Many papers have been published in support of each of
these opposing viewpoints. Some researchers insist that all
energetic particles arriving at Earth are accelerated exclusively
by CME-driven shocks, rather far from flare regions (Reames, 1999;
Kahler, 2001). Some others argue that traveling shocks could
perhaps contribute to lower-energy proton fluxes ($E < 10$ MeV),
while higher-energy ones ($E > 100$ MeV) are flare-accelerated on
the Sun (\textit{e.g.}, Klein and Trottet, 2001; Chertok, 1995;
Livshits and Belov, 2004; Li \textit{et al.}, 2007a,b).

The 20 January 2005 event provides a unique chance to address this
problem for the following reasons: (a)~availability of
spectral-imaging gamma-ray data from the Reuven Ramaty High Energy
Solar Spectroscopic Imager (RHESSI; Lin \textit{et al.}, 2002),
(b)~availability of highest-energy spectral gamma-ray data up to
200~MeV from the non-imaging SOlar Neutron and Gamma experiment
(SONG; Kuznetsov \textit{et al.}, 2004, 2008) on board the
CORONAS-F spacecraft (Oraevsky and Sobelman, 2002), (c)~the
strongest GLE in over half a century of observations, caused by
the unusually prompt arrival of protons at Earth (Bieber
\textit{et al.}, 2005; Belov \textit{et al.}, 2005; Simnett, 2006;
Cliver, 2006), and (d)~availability of many other kinds of
observational data, for both the flare region on the Sun, and for
the near-Earth space, up to the Earth's surface. In particular,
the source region on the Sun was observed in soft X-rays (SXR) by
GOES-10 and GOES-12 including the Soft X-ray Imager (SXI; Hill
\textit{et al.}, 2005). The H$\alpha$ flare was recorded at the
IPS Culgoora Solar Observatory in Australia and at the Hiraiso
Observatory in Japan. Nobeyama Radio Polarimeters (NoRP; Nakajima
\textit{et al.}, 1985; Torii \textit{et al.}, 1979) recorded an
unusually strong microwave burst at six frequencies: 2, 3.75, 9.4,
17, 35, and 80 GHz. The Transition Region and Coronal Explorer
(TRACE; Handy \textit{et al.}, 1999) observed the flare after its
peak in the 1600 \AA\ channel. The Extreme-Ultraviolet Imaging
Telescope (EIT; Delaboudini\`ere \textit{et al.}, 1995) and the
Large-Angle Spectrometric Coronagraph (LASCO; Brueckner \textit{et
al.}, 1995) on SOHO also supplied images related to this event.

In this paper, we first analyze in detail the flare itself. We
consider its source region on the Sun in different spectral
domains, using available imaging data and reconciling time
profiles of various emissions as well as spatial and spectral
characteristics of their sources. In this way, we endeavor to
understand this remarkable flare, its configuration, and the
physical conditions in the flare region. Then, based on the
results of this analysis, we consider which of the two
above-mentioned models of the origin of high-energy protons better
matches observations of the 20 January 2005 event. Our methodology
is based on joint analysis of multi-spectral data. To minimize the
influence of instrumental effects and model-dependent estimations,
we verify our results and estimates by using different methods and
independent observations as much as possible. Their agreement
enhances the reliability of our results and conclusions. Is is
important to analyze manifestations of accelerated protons
primarily during their main peak (corresponding to the main phase
of the highest rate of flare energy release), when the influence
of subsequent post-eruptive processes and transport effects on the
Sun and/or in interplanetary space, as well as probable
contributions from different acceleration mechanisms, which are
difficult to distinguish, is minimal.

We start our consideration in Section~\ref{Observations_Analysis}
with observational solar data and their brief analysis, including
the development of the active region and pre-flare situation.
Then, flare manifestations in gamma-rays, hard and soft X-rays,
microwaves, H$\alpha$, 1600~\AA, and extreme-ultraviolet (EUV)
emissions, as well as the CME are addressed. We conclude that the
extreme features of the flare were due to its occurrence in the
strong magnetic fields just above sunspot umbrae.
Section~\ref{SEP_GLE} is devoted to relating flare features to the
observed SEP/GLE characteristics. In particular, its proton
productivity is compared with other events from the same active
region. Finally, the temporal parameters of the impulsive parts of
the gamma-ray burst and GLE are compared. The same population of
energetic particles appears to be responsible for both the
$\pi^0$-decay gamma-ray burst and the leading spike of the GLE. In
turn, the $\pi^0$-decay emission was close, but not identical to,
emissions whose sources were located within the solar active
region. In Section~\ref{Discussion}, we discuss the results of our
analysis and argue that the $\pi^0$-decay emission also originated
in the flare region and that data available to us appear to favor
the flare-related initial acceleration of particles responsible
for the leading GLE spike, but do not preclude their origin in a
CME-driven shock. Section~\ref{Summary} summarizes the results of
our analyses of the flare and origin of SEP/GLE particles and
briefly addresses their implications.

\section{Observations and Analysis}
\label{Observations_Analysis}

The eruptive solar event of 20 January 2005 had several
outstanding properties. Many solar events are known, but only a
small number of them are comparable with this event in some of its
features. The high-energy near-Earth proton enhancement was
unprecedented. Since 1975, more than 600 proton events with
$E_\mathrm{P}>100\,\mathrm{MeV}$ have been registered; however,
high-energy proton fluxes exceeded 100~pfu in only five events. On
20 January 2005, the peak flux for
$E_\mathrm{P}>100\,\mathrm{MeV}$ protons reached the record value
of 650~pfu. At lower energies, the near-Earth proton flux was not
extremely intense, about 2000~pfu. A very strong, collimated
high-energy beam-like proton stream arrived at the Earth's orbit
almost simultaneously with electromagnetic flare emissions and
produced the GLE, which was the largest one in the last solar
cycle and one of the largest ever observed (Belov \textit{et al.},
2005; Simnett, 2006; Bieber \textit{et al.}, 2005; Cliver, 2006).
The largest effect, by an order of magnitude, was recorded with
antarctic neutron monitors.

Strong hard X-ray (HXR) and pronounced gamma-ray emissions up to
17 MeV were recorded with RHESSI (Krucker, Hurford, and Lin, 2005;
Hurford \textit{et al.}, 2006b; Share \textit{et al.}, 2006).
CORONAS-F/SONG detected gamma-rays of $>\,$100~MeV. Only a few
solar events are known in which such high-energy gamma-rays have
been recorded (\textit{e.g.}, Ryan, 2000).

The CME launched in this event was well observed with SOHO/LASCO
in only one C2 frame at $\approx 4.4 R_\odot$
(06:54\footnote{Times hereafter are UT, if otherwise not
specified}). Subsequent LASCO frames are severely contaminated
with traces of energetic particles; thus, the CME speed is rather
uncertain. Gopalswamy \textit{et al.} (2005) estimated it to be up
to 3242~km\thinspace s$^{-1}$, and the estimate of Simnett (2006)
was 2500 km\thinspace s$^{-1}$.

The microwave burst was huge, up to $\approx 10^5$ sfu, and was
observed up to 80 GHz. The associated long-duration flare was
strong, but not extraordinary,  both in soft X-ray and H$\alpha$
emissions (X7.1/2B). We analyze the properties of this superevent
and show some of them to be interrelated.

\subsection{History of the Active Region and Pre-Flare Situation}
\label{Sect_Magnetograms}

\begin{figure}    
   \centerline{\includegraphics[width=\textwidth,clip=]{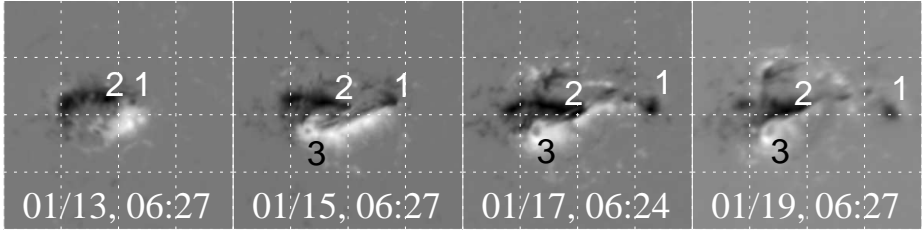}
              }
              \caption{
Evolution of AR 10720: SOHO/MDI magnetograms for 13\,--\,19 January.
Field of view is $295^{\prime \prime} \times 295^{\prime \prime}$.
Labels 1\,--\,3 mark features discussed in the text.
              }
 \label{magnetograms}
   \end{figure}

The event of 20 January 2005 occurred in active region 10720
(N14~W60). This region evolved rapidly. Its evolution is shown by
photospheric magnetograms observed with the Michelson Doppler
Imager (MDI; Scherrer \textit{et al.}, 1995) on SOHO.
Figure~\ref{magnetograms} presents a set of MDI magnetograms of AR
10720 from 13 through 19 January. Differential rotation is
compensated to 15 January, 00~UT. The magnetograms show two major
domains, the northern one of S-polarity (dark), and the southern
one of N-polarity (bright).

From 15 January onwards, the maximum magnetic field strength
exceeded the measurable capabilities of the MDI\footnote{often
referred to as ``high field saturation'', see
http:/\negthinspace/soi.stanford.edu/data/cal/}. This shows up as
a darker patch \textit{3} within the bright N-polarity domain. On
20 January, the active region was not far from the limb, leading
to significant distortions of the magnetograms. We therefore refer
to a magnetogram obtained on 18 January\footnote{We use MDI
magnetograms recalibrated in 2007}. On this day, the maximum
strength in a ``non-saturated'' part of the N-polarity area
reached $+3120$\thinspace G, while the persistence of patch
\textit{3} in this area suggests the magnetic field to be even
stronger there. The maximum strength in the S-polarity area was
$-2700$\thinspace G.

Noticeable during 13\,--\,20 January are (1) large shear motion
along the neutral line, clearly visible from the westward drift of
the westernmost S-polarity spot \textit{1}, and (2) the
displacement of the large northern S-polarity domain \textit{2}
southwards, approaching the N-domain, while the latter maintained
its latitude. This convergence brought the opposite-polarity areas
into contact on 17 January. These features of its development
suggest the high flare productivity of AR 10720 (cf.,
\textit{e.g.}, Birn \textit{et al.}, 2000). Indeed, it produced 17
M-class and 5 X-class events from 14 through 23 January. Many of
these were associated with CMEs.

\begin{figure}    
   \centerline{\includegraphics[width=\textwidth,clip=]{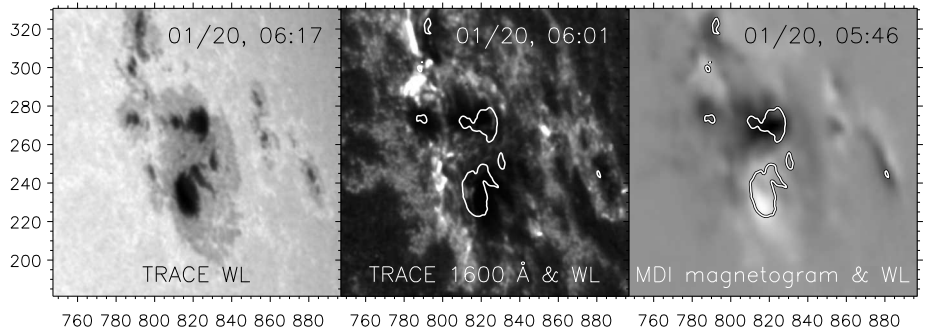}
              }
              \caption{
Pre-flare situation. TRACE WL image (left); TRACE 1600~\AA\ image
(middle) and SOHO/MDI magnetogram (right) overlaid with contours of
the sunspot umbrae visible in the WL image. All images are
``co-rotated'' to 06:52:30~UT. Axes in all images hereafter show the
distance in arcsec from the solar disk center.
              }
 \label{pre_flare}
   \end{figure}

Figure~\ref{pre_flare} presents the pre-flare situation in white
light (WL) and at 1600\thinspace \AA\ (TRACE), along with a
magnetogram. Coordinates are referred to RHESSI pointing. The
TRACE absolute pointing coordinates have an uncertainty larger
than its spatial resolution\footnote{see http:/\negthinspace
/trace.lmsal.com/Project/Instrument/cal/pointing.html \\ and
http:/\negthinspace/trace.lmsal.com/tag/}, whereas those of RHESSI
are believed to be accurate (and we checked this by comparing
TRACE WL and a full-disk MDI continuum image). The information
required to accurately coalign TRACE and RHESSI images was
obtained in the course of our analysis
(Section~\ref{TRACE_RHESSI_analysis}).

The area of the largest N-polarity sunspot's umbra is about 1.74
larger than the largest S-polarity one. All of these properties of
the sunspots imply that the real magnetic field is probably
stronger in the N-polarity umbra, while its total magnetic flux is
certainly larger.

\textit{The evolution of AR~10720 clearly points to a high flare
productivity, which indeed took place. Its magnetic fields reached
extreme strengths.} The active region became ``ready'' to produce
its strongest flare.

\subsection{Manifestations in the Lower Solar Atmosphere
and Hard Electromagnetic Emissions}

\subsubsection{Hard X-ray and Gamma-Ray Time Profiles}
\label{HXR_gamma_timeprofiles}

\begin{figure}    
   \centerline{\includegraphics[width=\textwidth,clip=]{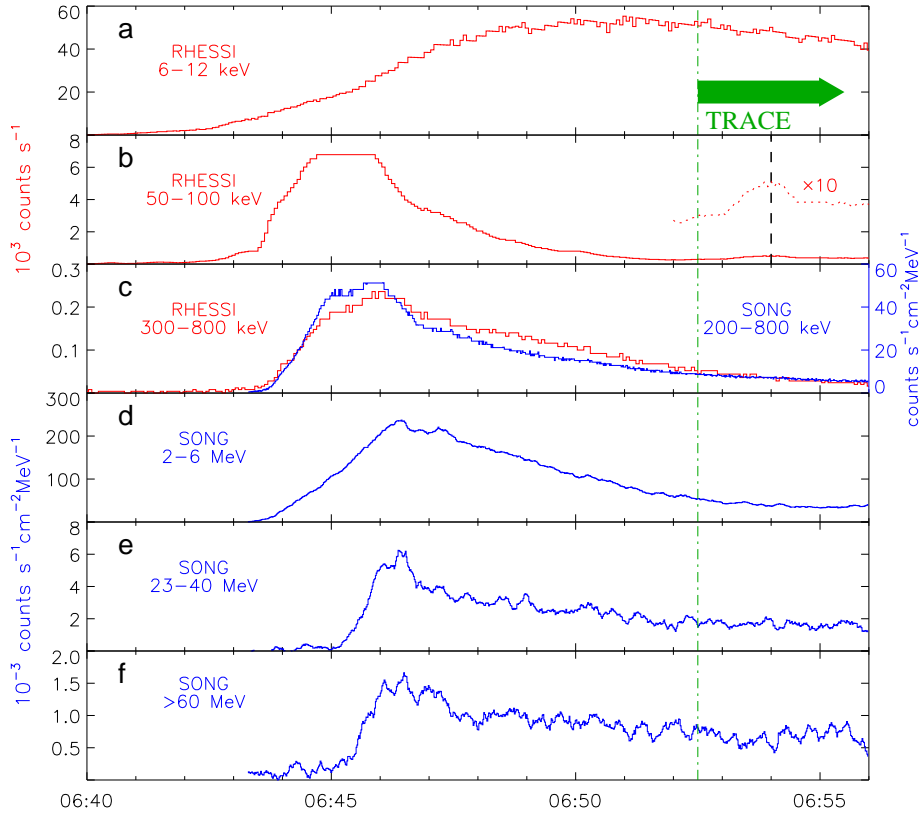}
              }
              \caption{
a\,--\,c: Some wide-band RHESSI time profiles (red). Green
dash-dotted line shows start of TRACE observations. Black dashed
line marks a weak enhancement of 25\,--\,300~keV emission. Red
dotted line shows part of its time profile enlarged ten-times.
c\,--\,f:~Some SONG time profiles smoothed over 13 points (blue;
the 200\,--\,800 keV channel has not been smoothed).
              }
 \label{RHESSI_SONG_timeprofs}
   \end{figure}

The event was observed in X-rays and gamma-rays by
RHESSI\footnote{RHESSI data used in this paper are preliminary.}
and CORON\-AS-F/SONG. Panels a\,--\,c of
Figure~\ref{RHESSI_SONG_timeprofs} show some of the wide-band
RHESSI time profiles (four second sampling rate) produced using
RHESSI Data Analysis Software. The time profiles in a range of
3\,--\,25~keV (\textit{e.g.}, panel \textit{a}) are similar and
smooth, with their maxima significantly delayed with respect to
those of higher energies, most likely due to the Neupert effect,
which we discuss later. Harder emissions became strong at
06:42:40, while softer X-ray emissions started earlier, probably
due to initial heating. The time profiles in a range of
25\,--\,300~keV are similar. They seem saturated around the flare
peak (this effect can be compensated for, see
Section~\ref{H_alpha_WL_HXR}). The highest-energy time profiles
(not shown) lag behind those of 25\,--\,300~keV and rise sharply
after 06:45.

SONG observed hard emissions in 12 channels from 43 keV to 310 MeV
with a one second sampling rate
(Figure~\ref{RHESSI_SONG_timeprofs}c\,--\,f). Its net count rates
were computed by subtracting background trends inferred from
previous orbits. The gamma-ray burst was reliably recorded from
06:43:30 to 06:58:30, then uncertainties become large due to
contamination from charged particles reaching the spacecraft. The
43\,--\,230 keV channels suffer from register overflow. Flux
variations recorded with RHESSI and SONG in corresponding energy
bands are close, any dissimilarities are probably due to
differences in energy bands. The differences in count rates
(\textit{e.g.}, panel \textit{c}) are due to different detectors.
Two peaks at about 06:45 and 06:46 are detectable in the records
of both instruments. The hardest emissions rise sharply after
06:45:30.

The time profiles in 50\,--\,100~keV and 200\,--\,800~keV bands
are largely similar. These time profiles and the highest-energy
($> 60\,$MeV) emissions have some similarities as well as
differences. The highest-energy gamma-rays start later than
lower-energy emissions, rise sharper, and decay slower. On the
other hand, subsidiary peaks in the $> 60\,$MeV band at 06:46,
06:46:30, and 06:47:15 have pronounced counterparts in
lower-energy bands. They are also detectable in
intermediate-energy bands, which are not shown.

\subsubsection{H$\alpha$, White Light, and Hard X-ray Images}
\label{H_alpha_WL_HXR}

\begin{figure}    
   \centerline{\includegraphics[width=\textwidth,clip=]{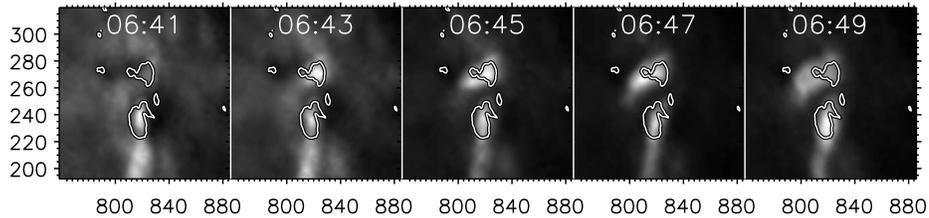}
              }
              \caption{
Development of flare ribbons in H$\alpha$ emission. Contours show
the sunspot umbrae.
              }
 \label{H_alpha_flare}
   \end{figure}

Culgoora Solar Observatory observed the H$\alpha$ flare every 20
seconds from 06:40:47 to 08:04. Figure~\ref{H_alpha_flare} shows
some of these images overlaid with contours of the sunspot umbrae
from the TRACE WL image (see Figure~\ref{pre_flare}). One can see
the development of H$\alpha$ flare ribbons with respect to
sunspots starting from the onset of the HXR emission up to its
cessation (maximum of SXR emission). Initially, H$\alpha$ emission
comes from the future southern ribbon, being concentrated in the
sunspot umbra. The emission south of this sunspot is due to weaker
initial flaring. At 06:43, when HXR emission becomes strong,
H$\alpha$ emission appears in the northern sunspot umbra. Later
on, the northern ribbon was significantly extended toward the
southeast.

\begin{figure}    
   \centerline{\includegraphics[width=0.7\textwidth,clip=]{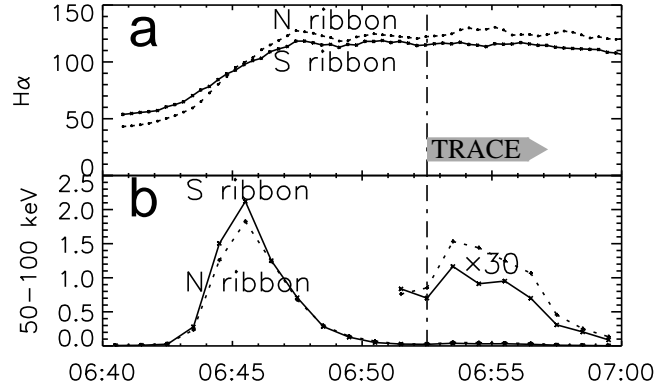}
              }
              \caption{
Time profiles of the average brightness (arbitrary units) computed
separately for northern (dotted) and southern (solid) ribbons:
H$\alpha$ (a, Culgoora Solar Observatory), 50\,--\,100~keV hard
X-rays (b, RHESSI). Dash-dotted line: start of TRACE observations.
Enlarged time profiles show a weak enhancement of the HXR emission
around 06:55.
              }
 \label{H_a_HXR_timeprofs}
   \end{figure}

The limited resolution of the H$\alpha$ images and atmospheric
effects do not permit consideration of bright kernels, but it is
possible to study larger features. Figure~\ref{H_a_HXR_timeprofs}a
shows time profiles of the total H$\alpha$ emission computed
separately for each ribbon. Their similarity suggests a single
common source for their emissions.

\begin{figure}    
   \centerline{\includegraphics[width=\textwidth,clip=]{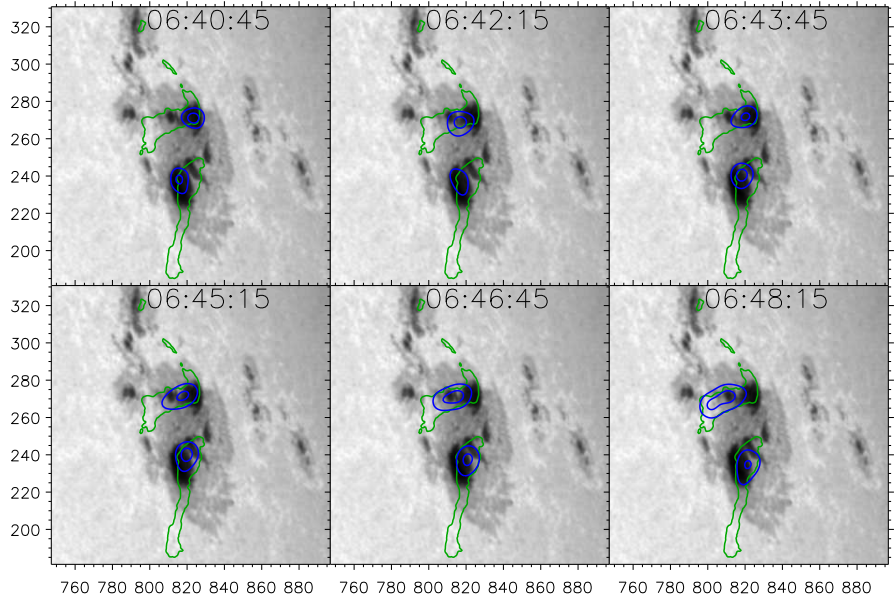}
              }
              \caption{
50\,--\,100 keV hard X-ray sources (RHESSI, integration time 30
seconds, blue contours) overlying the TRACE WL image (gray scale).
Green contours delineate flare ribbons in TRACE 1600~\AA\ image at
06:52:30. All images are referred to RHESSI pointing.
              }
 \label{HXR_images}
   \end{figure}

Figure~\ref{HXR_images} shows the relative positions of 50\,--\,100~keV
HXR sources, sunspots, and flare ribbons around the time of the main flare
phase. At 06:40:45, both HXR sources are located in the sunspot
umbrae. The active northern source shifts firstly east along the
ribbon, and then, when HXR flux sharply increases, it returns to the
umbra. Then, it extends along the ribbon, and during the maximum
phase moves to the east. The southern source is much steadier.

HXR sources are located on the H$\alpha$ ribbons. Similar
development of the flare ribbons and HXR sources is also shown by
a movie of this
event\footnote{http:/\negthinspace/svs.gsfc.nasa.gov/vis/a000000/a003100/a003162/},
TRACE 1600\thinspace \AA\ + RHESSI/SXR \& HXR. H$\alpha$ and HXR
images show the following features: (a)~both ribbons cross the
umbrae of sunspots, (b)~initially, HXR and H$\alpha$ emission
sources are located in the sunspot umbrae, and then, after 06:45,
the northern source shifts from the umbra to a region of weaker
magnetic fields; \textit{the gamma-ray emission maximum occurs
just at this time}, and (c)~the displacements of the southern HXR
and H$\alpha$ sources are relatively small.

To compute spatially-resolved HXR time profiles, we produced
50\,--\,100~keV RHESSI images using the Pixon method.
Figure~\ref{H_a_HXR_timeprofs}b shows HXR time profiles computed
separately for each ribbon. Again, the time profiles of both ribbons
are so similar as to suggest a common source for their emissions.
The H$\alpha$ time profiles are much smoother than those of
HXR and lag behind them due to the well-known long decay time of
H$\alpha$ emission.

\subsubsection{1600 \AA\ and Hard X-Ray Images}
\label{TRACE_RHESSI_analysis}

The TRACE 1600~\AA\ channel is sensitive to emissions from
4\,000\,--\,10\,000~K plasmas; hence the brightest features in
H$\alpha$ and 1600~\AA\ images show basically the same structures,
\textit{i.e.}, footpoint regions of flaring coronal loops and,
sometimes, the cooled loops themselves. Precipitation of
high-energy electrons in the chromosphere causes its local
heating, HXR bremsstrahl\-ung, and the excitation of H$\alpha$ and
1600~\AA\ emissions from lower layers of the solar atmosphere.
Therefore, a correlation is expected between the positions and
time profiles of HXR and 1600~\AA\ sources. TRACE had a break in
observations between 06:18 and 06:52:30; afterwards, it showed
well-developed flare ribbons.

Using variance and correlation analyses (Grechnev, 2003; Slemzin,
Kuzin, and Bogachev, 2005) of TRACE 1600~\AA\ images observed
during 06:52:41\,--\,06:57:30, we revealed several pairs of
kernels with variable brightness (cf. Su \textit{et al.}, 2006).
The paired kernels, located in different ribbons with similar time
profiles, suggest a common source of precipitated electrons. They
most likely represent conjugate footpoints of the loops connecting
them (cf. Grechnev, Kundu, and Nindos, 2006). Unlike the main
phase of the flare, the kernels were located at the outer sides of
the ribbons, far from the sunspot's umbrae. This outward motion
corresponds to the expansion of the ribbons.

We then performed the variance and correlation analyses for the
RHESSI 50\,--\,100 keV images produced using the CLEAN method over
the same time interval, when the HXR emission was enhanced (see
Figures~\ref{RHESSI_SONG_timeprofs}, \ref{H_a_HXR_timeprofs}). It
revealed regions with a configuration corresponding closely with
the TRACE 1600~\AA\ kernels, but displaced by $[+7.5^{\prime
\prime}, -10.5^{\prime \prime}]$. This provided us with the
information required to coalign TRACE and RHESSI images with an
accuracy of the order of $\approx 1^{\prime \prime}$. This also
confirms that the ribbons represent the bases of closed loops.

As established in
Sections~\ref{HXR_gamma_timeprofiles}\,--\,\ref{TRACE_RHESSI_analysis},
\textit{atypically, the flare ribbons in this event crossed the
sunspot umbrae. The coronal configuration above the ribbons was
closed. During the main phase, the brightest H$\alpha$ and HXR
emissions came from the sunspot umbrae. The flare occurred in very
strong magnetic fields.}

\subsubsection{HXR and Gamma-Ray Spectra and Images}

\begin{figure}    
   \centerline{\includegraphics[width=\textwidth,clip=]{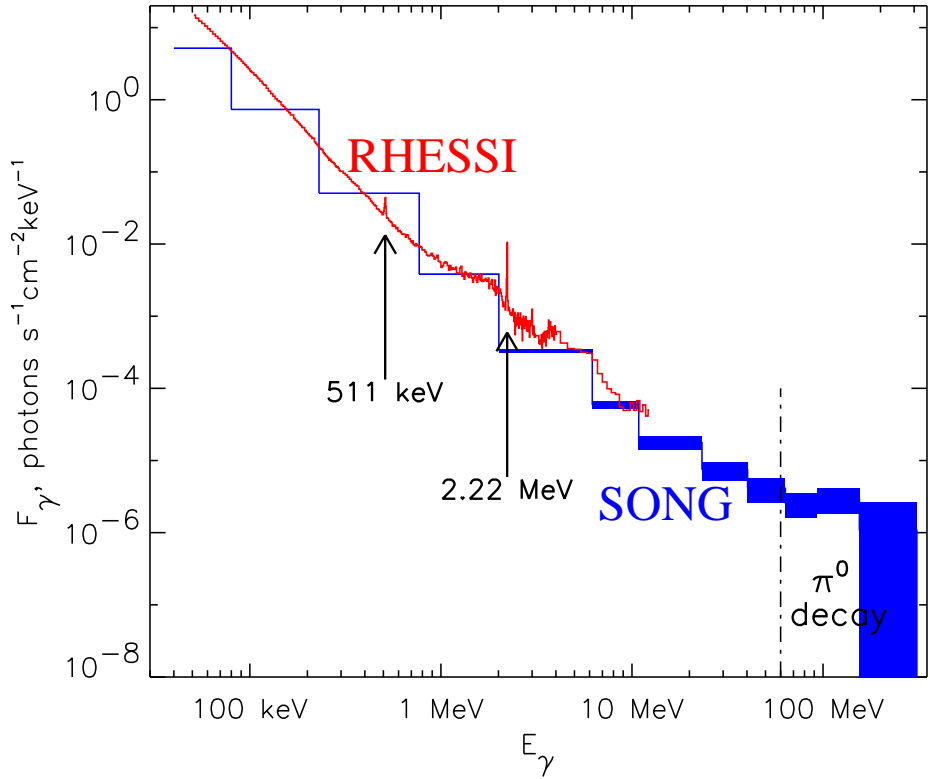}
              }
              \caption{
High-resolution RHESSI (red) and wide-range SONG (blue) spectra
recorded during the hardest flare emission. Vertical dash-dotted
line shows the high-energy part of the SONG spectrum where the
$\pi^0$-decay emission dominates. Shading shows uncertainties.
              }
 \label{X-ray_spectrum}
   \end{figure}

As Figure~\ref{RHESSI_SONG_timeprofs} shows, the spectrum became
harder after 06:45~UT. Figure~\ref{X-ray_spectrum} presents the
non-thermal parts of the spectra recorded with RHESSI and SONG
just after the flare peak. The detailed RHESSI spectrum was
computed using the OSPEX software. The SONG spectrum (Kuznetsov
\textit{et al.}, 2007) extends up to the highest energies. Data
from both instruments largely agree with each other
quantitatively; however, the lowest SONG band suffers from
overflow, while the absolute calibration of highest-energy RHESSI
data is preliminary. One can distinguish several components in
both spectra. These components correspond to the expected
electron-bremsstrahlung and nuclear gamma-ray features.

The power-law index of bremsstrahlung is $\gamma = 2.8-2.9$, which
corresponds to an electron index of $\delta = 4.4-4.5$ under
thick-target conditions.

The RHESSI spectrum shows a prompt electron-positron annihilation
line of 511 keV and a delayed neutron-capture line of 2.22 MeV.
The most intense is the discrete 2.22 MeV line, which is emitted
in neutron capture on hydrogen. Neutrons appear in nuclear
reactions of accelerated ions, streaming down to the photosphere,
with its material. Neutrons undergo thermalization in $\approx$80
seconds, and some of them are eventually captured on hydrogen,
$\mathrm{H} + n \to 2\mathrm{H} + \gamma$, with the emission of a
2.22 MeV photon. Thus, the discrete 2.22 MeV line is indicative of
accelerated ions with an energy exceeding a few MeV.

The SONG spectrum reveals an enhancement at the highest energies,
$E_\gamma > 60$~MeV, due to Doppler-broadened emission of $\pi^0$
decay into two photons (Kuznetsov \textit{et al.}, 2007). The
neutral pions appear in proton-proton collisions when
$E_\mathrm{P} > 300$ MeV (Ramaty, Kozlovsky, and Lingelfelter,
1975; Hudson and Ryan, 1995). The $\pi^0$-decay emission leads the
2.22 MeV line emission by about 80 seconds (see
Figure~\ref{GLE_gamma}), which is consistent with the idea that
the populations of incident protons responsible for these
emissions were close.

The discrete 2.22 MeV line emission, which is indicative of the
precipitation of accelerated protons and heavier ions, is strong
enough to allow one to construct an image of its source. We
computed it from RHESSI data using the Pixon method. To compare it
with the bremsstrahlung component, we also produced a RHESSI image
in a band of 100\,--\,300~keV for the interval of
06:46:00\,--\,06:46:30 using the Pixon method. This imaging
interval corresponds to strong high-energy gamma-ray emission.
Positions and configurations of HXR sources at 50\,--\,100 keV and
100\,--\,300 keV are close for as long as they are detectable (see
Hurford \textit{et al.}, 2006b; Share \textit{et al.}, 2006).

\begin{figure}    
   \centerline{\includegraphics[width=\textwidth,clip=]{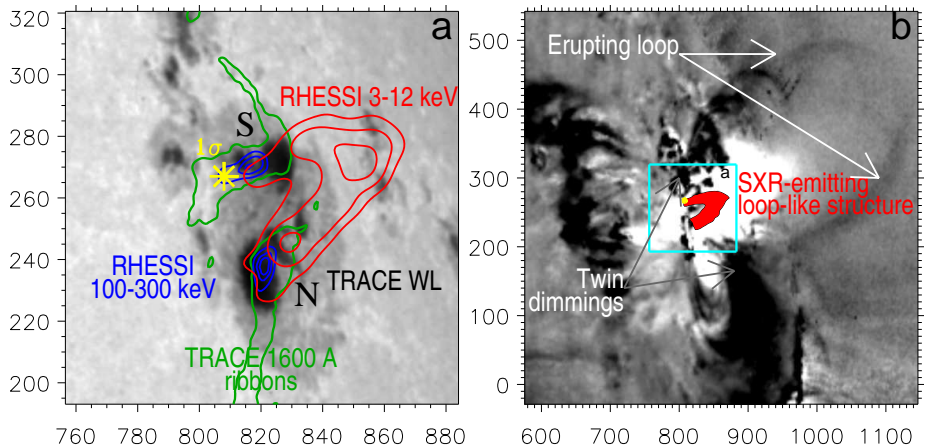}
              }
              \caption{
(a)~Contours of RHESSI HXR, SXR, and gamma-ray images overlying
the TRACE WL image at 06:00 shown as gray scale background. Yellow
star: 2.22~MeV line centroid (1$\sigma$ error size), blue
contours: 100\,--\,300~keV image at 06:46:00\,--\,06:46:30, [20,
50, 80]\%; red contours: 3\,--\,25 keV, 06:41:00\,--\,06:48:30,
[30, 50, 80]\%; green contours: flare ribbons from TRACE 1600~\AA\
image at 06:52:30. ``N'' and ``S'' denote polarities of the main
magnetic domains. (b)~Dimmings revealed from EIT 195~\AA\ images
and the loop-like structure (red) visible in the EIT~171~\AA\
band. Turquoise frame: field of view of panel (a).
              }
 \label{RHESSI_gamma_images}
   \end{figure}

Figure~\ref{RHESSI_gamma_images}a presents the RHESSI 100\,--\,300
keV image (blue contours), along with green contours indicating
flare ribbons (from TRACE 1600~\AA\ image at 06:52:30~UT), on top
of the TRACE WL image shown as halftone background. A large 2.22
MeV line source (beam size $\approx 35^{\prime \prime}$) is
located within the active region with its centroid (yellow star)
being in the northern ribbon. The positions of the sources in
Figure~\ref{RHESSI_gamma_images}a agree with those presented by
Hurford \textit{et al.} (2006b).

When gamma-ray emission reached its maximum, the northern HXR
source became extended, with its eastern part coincident with the
2.22 MeV source. Independent of instrumental limitations and
collimation of particle streams, the fact that the 2.22 MeV line
source centroid was aligned with a flare ribbon means that
\textit{accelerated ions precipitated in the region, where coronal
structures were closed at that time}.

\subsection{Coronal Manifestations and Soft X-Ray Emission}
\label{Coronal_structure}

\subsubsection{Soft X-Ray RHESSI, GOES/SXI, and SOHO/EIT Images}

A large loop-like SXR-emitting structure with its top reaching a
height of $\approx 40\,000$ km and legs rooted in both sunspot
umbrae is observed with EIT in the 171~\AA\ channel, RHESSI in the
3\,--\,25~keV bands (Figure~\ref{RHESSI_gamma_images}a,b), and
GOES/SXI. The fact that these instruments each show a similar
structure confirms its coronal nature and predominantly thermal
emission. The coronal structure looks like a single thick loop
with a blob-like enhancement at its top, or above it, visible in
RHESSI images produced using the CLEAN method. To study this blob
in detail is beyond the scope of this paper. The loop-like
structure is hot, as RHESSI and GOES/SXI show, and probably
consists of a bundle of thinner SXR-emitting loops. At least half
an hour before the flare maximum, RHESSI detected the blob at its
top (a gradual increase of SXR flux started at 06:00 according to
GOES-10 and GOES-12), and then its legs became detectable. Around
the flare peak in soft X-rays, the emission from the whole Sun was
dominated by this large loop-like structure. It was detectable in
SXI images up to at least 08~UT and then its emission gradually
declined.

To estimate its parameters, we produced a spectrum from RHESSI
averaged data over 06:46\,--\,06:47 using OSPEX software. Then, we
fitted the spectrum with a single thermal component dominating up
to 25\,--\,30~keV, and a single power-law component. These
estimations provide a temperature of $T_\mathrm{e} \approx\,$32~MK
and a total emission measure of EM$ \approx 7.6 \times
10^{49}$~cm$^{-3}$.

After 07 UT, an apparently expanding arcade appears in 1600~\AA,
H$\alpha$, and EUV images well below the large SXR-emitting
structure. The loops of this arcade are detectable in SOHO/EIT and
GOES/SXI images up to the end of the day. This arcade does not
seem to be directly associated with the SXR-emitting loop-like
structure. The paired kernels revealed from the analysis of HXR
and 1600~\AA\ images during the enhancement around 06:55, well
after the main flare peak, do not coincide with the footpoints of
the loop-like structure and most likely show the formation of the
arcade underneath.

From 06:54 onwards, SOHO/EIT and LASCO frames become increasingly
contaminated by the traces of high-energy particles impacting the
CCD detector (cf. Andrews, 2001), and after 07:14 the images
become indecipherable. This effect indicates that protons with
$E_\mathrm{P} > 40$ MeV and up to at least 400~MeV reached SOHO
(Didkovsky \textit{et al.}, 2006). Nevertheless, it is possible to
reveal the structure of the coronal dimming using several frames
(Grechnev, 2004; Chertok and Grechnev, 2005).
Figure~\ref{RHESSI_gamma_images}b shows a dimming ``portrait''
composed of minima over each pixel in the EIT 195~\AA\ fixed-base
(06:24:10) difference frames from 06:36 through 09:19~UT (this EIT
channel has a temperature sensitivity maximum at 1.3~MK). Dimmings
are seen as rather large, dark areas, while the red-filled contour
shows a bright, overexposed image of the loop-like structure
recorded in the EIT 171~\AA\ channel. ``Twin'' dimmings are
located near the ends of the post-eruptive arcade (here its image
is inside a bright area) northwest and southwest of the ribbons
and the large loop-like structure (cf.
Figure~\ref{RHESSI_gamma_images}a). The 2.22 MeV line source, like
the other major flare sources, is located far from the dimming
regions.

\subsubsection{Soft X-Ray GOES Light Curves and the Neupert Effect}

\begin{figure}    
   \centerline{\includegraphics[width=\textwidth,clip=]{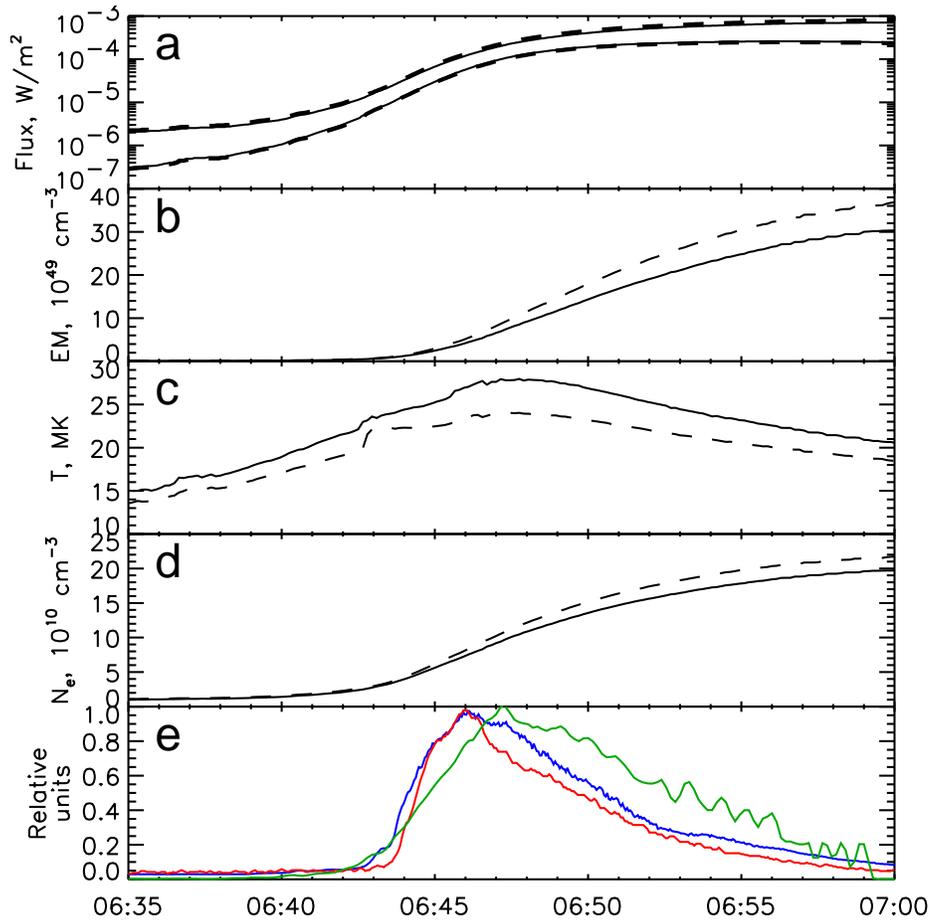}
              }
              \caption{
GOES-10 (dashed) and GOES-12 (solid) soft X-ray flux density (a),
emission measure (b), temperature (c), and density (d). (e)~Neupert
effect. Normalized time profile of the 300\,--\,800~keV (red)
emission in comparison with microwaves at 35~GHz (blue) and a
derivative of the soft X-ray flux at 1\,--\,8~\AA\ (GOES-10, green).
              }
 \label{GOES_data}
   \end{figure}

Figure~\ref{GOES_data} shows in panel (a) soft X-ray light curves
recorded with GOES-10 (dashed) and GOES-12 (solid). We have
computed a total emission measure EM (b) and an average
temperature $T_{\mathrm e}$ (c) from two GOES bands (White,
Thomas, and Schwartz, 2005). These values over 06:46\,--\,06:47
are close to those computed from RHESSI data for the same time
interval. Because the large loop-like structure alone produced
practically the whole solar SXR emission around the flare peak, we
estimate the plasma density in this structure
(Figure~\ref{GOES_data}d) taking its parameters from SXR images,
\textit{i.e.}, width of a leg $\approx 9\times 10^8$~cm, footpoint
separation $\approx 2.5\times 10^9$ cm, height $\approx 4\times
10^9$ cm, volume $\approx 3.5\times 10^{27}$ cm$^3$, and a visible
area of the whole loop, $4\times 10^{18}$ cm$^2$.

Figure~\ref{GOES_data}e shows a comparison of the unsaturated
300\,--\,800 keV RHESSI time profile with a temporal derivative of
the SXR flux recorded in the GOES-10 1\,--\,8~\AA\ channel. Their
similarity shows the Neupert effect (Neupert 1968), suggesting
that the SXR-emitting plasmas ``evaporated'' from the chromosphere
due to energy lost by precipitating high-energy particles, which
produced the observed HXR emission (see also Liu \textit{et al.},
2006). Since the whole SXR flux came from the loop-like structure
with its footpoint regions coinciding with HXR sources, and the
Neupert effect proves the connection between HXR and SXR
emissions, we conclude that the HXR emission in this phase was
produced by high-energy electrons precipitating from this
loop-like structure. Heated and ``evaporated'' plasmas, in turn,
re-entered this structure. This also confirms that flare processes
occurred within a closed configuration.

\subsection{Microwave/Millimeter Burst}
\label{MW_burst}

\begin{figure}    
   \centerline{\includegraphics[width=\textwidth,clip=]{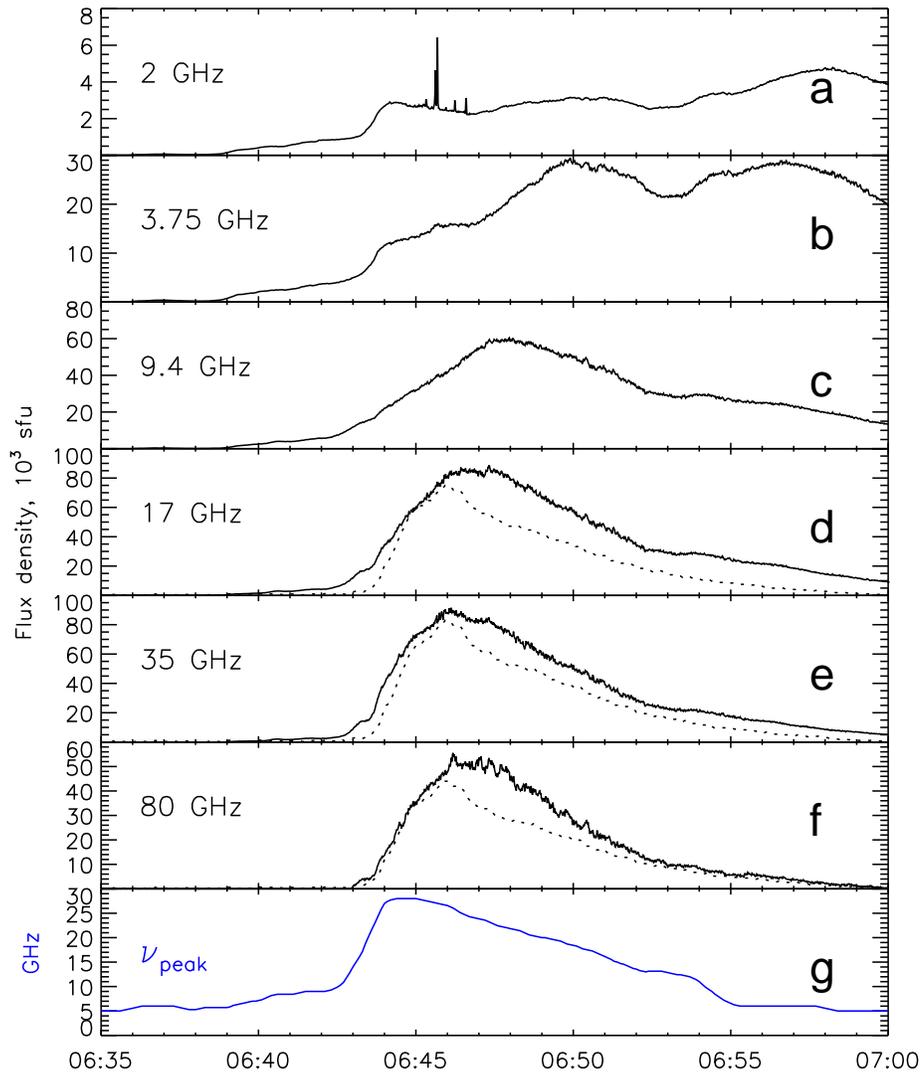}
              }
              \caption{
(a\,--\,f)~NoRP total flux time profiles at 2\,--\,80 GHz, Stokes
$I$. The vertical axes are in units of $10^3$~sfu. Dotted lines
show the hard X-ray emission at 300\,--\,800 keV according to
RHESSI. (g)~The turnover frequency of the microwave total flux
spectrum according to NoRP data.
              }
 \label{norp_timeprofs}
   \end{figure}

Total flux records from NoRP (Figure~\ref{norp_timeprofs}a\,--\,f)
show a huge radio burst with its flux density reaching almost
$10^5$~sfu. A strong emission up to $5\times 10^4$~sfu was
recorded even at 80 GHz. The 80~GHz data had calibration problems
(up to June 2005), which we fixed by means of a time-dependent
correction inferred using hard copies of records from the detector
of the radiometer. The time profiles at 17, 35, 80~GHz, and at
300\,--\,800 keV (dotted line in
Figure~\ref{norp_timeprofs}d\,--\,f) are similar, thus suggesting
a closeness of the sources responsible for radio and hard
electromagnetic emissions. Until the onset of the $\pi^0$-decay
emission at 06:45, features detectable at most frequencies
correspond. The time profiles over 17\,--\,80 GHz start to show
more fluctuations at this time (about 06:52). The microwave flux
densities at these frequencies reach their maxima within the
interval of 06:46\,--\,06:47, which corresponds to the maximum of
the hardest gamma-rays emissions.

\begin{figure}    
   \centerline{\includegraphics[width=\textwidth,clip=]{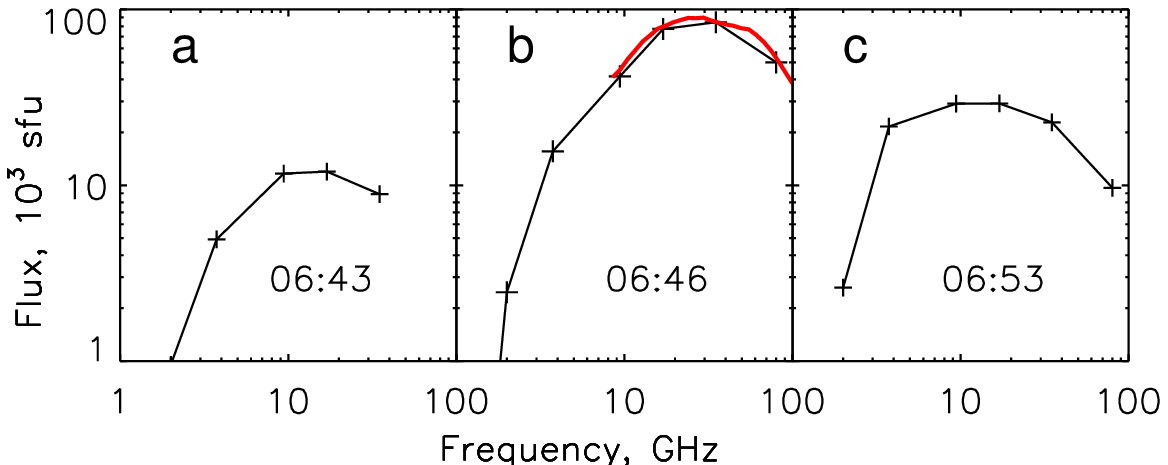}
              }
              \caption{
Total flux radio spectra computed from NoRP data (Stokes $I$, $10^3$
sfu) before (a), during (b), and after the flare peak (c). Red:
spectrum modeled with Ramaty code.
              }
 \label{NoRP_spec}
   \end{figure}

Figure~\ref{NoRP_spec} presents radio spectra computed from the
NoRP data before, during, and after the flare maximum, with each
point averaged over 5 seconds. The turnover frequency
$\nu_\mathrm{peak}$ computed using interpolation is shown in
Figure~\ref{norp_timeprofs}g. It is $\approx 30$~GHz around the
peak time, suggesting that microwaves came mainly from the
strongest-field regions above the sunspots' umbrae.

The microwave spectral index computed from the ratios of fluxes at
the highest frequencies is very hard, $\alpha \lsim 1$. The spectra
imply, however, that the slope at frequencies $>$80~GHz could be
steeper, and real $\alpha$ larger. The electron power-law index
corresponding to the optically thin limit is $\delta \approx 2.5$
(Dulk, 1985). From the HXR spectrum we found $\delta \approx
4.4-4.5$. The trapping effect hardens the microwave-emitting
electron spectrum up to $\delta - 2$ (Silva, Wang, and Gary, 2000).
Hence, a plausible power-law index of microwave-emitting electrons
on 20 January 2005 was $\delta \approx 2.5-3.5$.

Near the peak of the burst, the density inferred from GOES data is
$n_0 \approx 7\times 10^{10}$~cm$^{-3}$. With an emitting region
depth of $2\times 10^8$~cm and the usually assumed relative number
of accelerated electrons $n_\mathrm{acc}/n_0 = 10^{-4}$, one
estimates a magnetic field strength of $B \approx 1500-2200$~G
(Dulk, 1985). Thus, a large number of hard-spectrum electrons were
gyrating in very strong magnetic fields. Since two sources in
adjacent footpoints of a loop emit, the turnover frequency in the
stronger-field footpoint was still higher.

Modeling the turnover and higher-frequency parts of the radio
spectrum observed near the flare peak using Ramaty code (Ramaty,
1969; Ramaty \textit{et al.}, 1994), provides a satisfactory fit
(red in Figure~\ref{NoRP_spec}b) with radio sources assumed to
correspond to the observed HXR sources, $B \approx 1600$~G in the
strongest-field region [($-2700$~G, $>3100$~G) at the
photosphere], $\delta \approx 2.5-3.5$, and the relative number of
accelerated electrons of $n_{(E
> 10\,\mathrm{keV})}/n_0 \approx 10^{-3}-10^{-2}$. The lower-frequency
microwaves were probably emitted by larger regions, as the spectra
in Figure~\ref{NoRP_spec} imply. The major conclusion is that
\textit{the observed long-millimeter emissions could originate
only in very strong magnetic fields just above sunspot umbrae, in
accordance with other data}.

\subsection{The CME and its Speed}
\label{CME_Speed}

\begin{figure}    
   \centerline{\includegraphics[width=\textwidth,clip=]{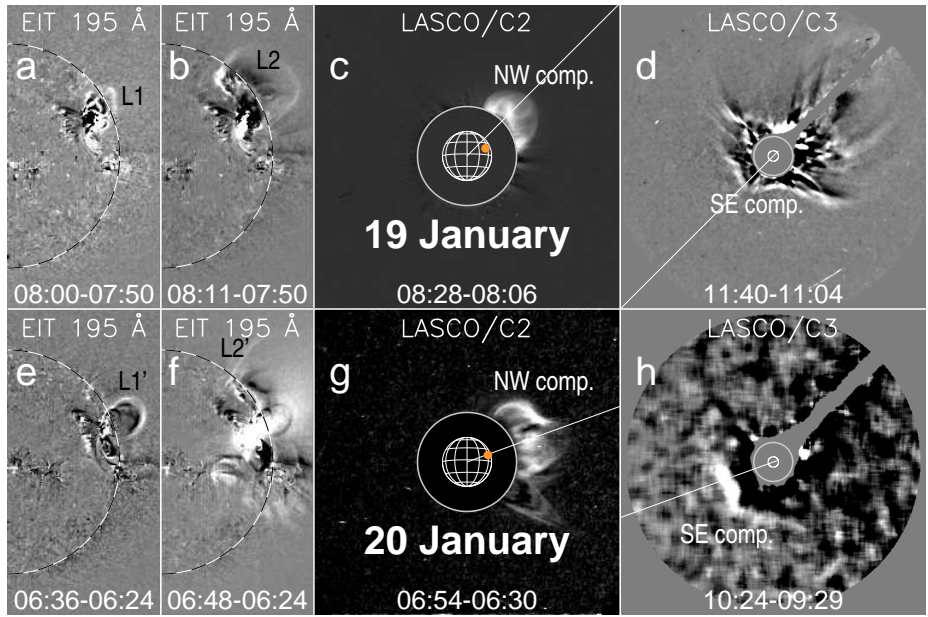}
              }
              \caption{
Homologous two-fold eruptions on 19 (top) and 20 (bottom) January
shown by EIT 195~\AA\ and LASCO/C2 \& C3 difference images. Small
orange disks denote AR~10720. Straight lines mark central position
angles of the CMEs. Dashed circles in EIT images and thin white
circles in LASCO images mark the limb. Larger gray circles denote
the occulting disks. LASCO images of 20 January were heavily
processed to reveal the CME.
              }
 \label{CMEs}
   \end{figure}

Estimations of the CME speed in this event are difficult due to
the prompt arrival at SOHO of a strong flux of high-energy
protons, whose impacts on the CCD detector severely contaminated
LASCO images. There is only one C2 frame at 06:54
(Figure~\ref{CMEs}g) in which the CME, with a central position
angle (PA) of $290^\circ-293^\circ$, is clearly detectable at
$\approx 4.4 R_\odot$. Assuming the loop-like structure, rising up
to $\approx 1.3R_{\odot}$ in the EIT 195~\AA\ frame at 06:36,
corresponds to the frontal structure of the CME, one estimates its
sky-plane speed $V\approx\,$2000 km\thinspace s$^{-1}$. From
similar considerations Gopalswamy \textit{et al.} (2005) and
Simnett (2006) obtained the speeds of 2100 and 2500 km\thinspace
s$^{-1}$.

Another evaluation of Gopalswamy \textit{et al.} (2005), that gave
a record speed of 3242~km\thinspace s$^{-1}$, seems to be greatly
overestimated. Combining the EIT data point at 06:36 with the
LASCO one in a second-order fit implies a constant CME
acceleration, although the acceleration is know to initially
increase sharply, from zero to a very high value, and then to
decline to zero (see, \textit{e.g.}, the review by D{\'e}moulin
and Vial, 1992; Uralov, Grechnev, and Hudson, 2005). The CME may
even decelerate. Moreover, as Zhang and Dere (2006) found, the
greater the acceleration is, the shorter the time it acts over.

The identification of structures seen in LASCO and EIT images
might be incorrect, especially the EIT loop and LASCO front.
Frontal structures are rarely observed in EIT images as faint
features indiscernible in non-subtracted frames (\textit{e.g.},
Dere \textit{et al.}, 1997; Uralov, Grechnev, and Hudson, 2005),
whereas this expanding loop is distinct in raw EIT images. Frontal
structures outrun all the other observable parts of the CME; thus,
combining the LASCO and EIT data points one can greatly
overestimate the CME speed.

However, the CME speed can be independently estimated from the
conspicuous resemblance of halo CMEs in four homologous major
events of 15\,--\,20 January. Each of them consisted of a bright
structure rapidly expanding in a NW direction and a fainter,
slower SE component. The speed ratios of CMEs on 15, 17, and 19
January were close, $\approx 3-4$. NW and SE components were also
present in the 20 January CME (Figure~\ref{CMEs}), and their
shapes were similar to those of preceding CMEs. As discussed
below, the speeds of the SE components of the 19-th
(360~km\thinspace s$^{-1}$) and 20-th (485~km\thinspace s$^{-1}$)
January CMEs were of the same order.

Figure~\ref{CMEs} shows this for the 19 and 20 January events. The
onsets of both events also exhibited a strong resemblance:
eruptions started with expansions of similar loops \textit{L1}
(\textit{L1}$^\prime$) visible in EIT images, and then much larger
similar loops \textit{L2} (\textit{L2}$^\prime$) started to
expand. There is no perfect analogy between frames \textit{a},
\textit{b} and \textit{e}, \textit{f} due to the meager imaging
rate of EIT, but the reordered sequence \textit{a,e,b,f} looks
impressive. Also very similar are disturbed regions eastward and
southward of AR~10720, bright chains between NE bases of loops
\textit{L1}, \textit{L2} (\textit{L1}$^\prime$,
\textit{L2}$^\prime$), and dimmings above the NW limb.

\begin{figure}    
   \centerline{\includegraphics[width=\textwidth,clip=]{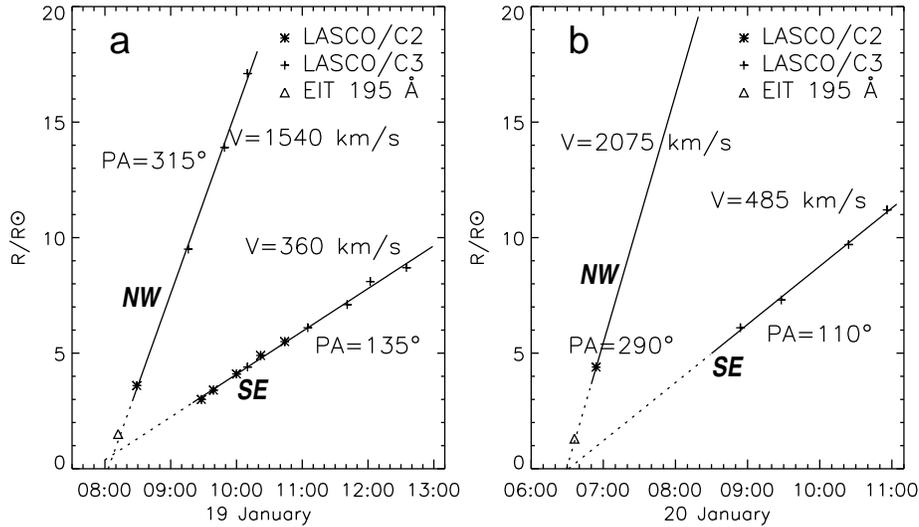}
              }
              \caption{
Height-time plots for NW and SE parts of the 19 (a) and 20 (b)
January CMEs.
              }
 \label{CME_speeds}
   \end{figure}

Let us compare the sky-plane speeds for the NW and SE parts of the
two CMEs. Their source regions were close, N15\thinspace W51 on 19
and N14\thinspace W61 on 20 January. In both cases, we measure the
speeds at the central position angles of the main NW parts. Since
the NW part of the 20 January CME is in question, we analyze its
analog on 19 January by considering its bright, sharp front rather
than its weaker leading aureole, as done in the SOHO LASCO CME
Catalog\footnote{http:/\negthinspace/cdaw.gsfc.nasa.gov/CME\_list/}.
As shown in Figure~\ref{CME_speeds}a, the NW
(PA$\approx$315$^\circ$) and SE ($\approx$135$^\circ$) components
of the 19 January CME had speeds of 1540 and 360~km\thinspace
s$^{-1}$. In four frames of a LASCO/C3 running-difference movie
for 20 January (08:54\,--\,10:56), despite strong contamination
from energetic particles, the SE edge at PA$\approx$110$^\circ$
can be detected moving away from the Sun. It is almost
indiscernible; hence, its measurements have poor accuracy. The
best frame of 10:24\,--\,09:29 in Figure~\ref{CMEs}h required
heavy processing to reveal. We estimate its speed as $\approx$485
km\thinspace s$^{-1}$. Simnett (2006) considered it as another CME
(with a speed of 604~km\thinspace s$^{-1}$), because its
back-projected time to cross the limb was 07:15, well after the
flare. However, the CME ejected from AR~10720, or nearby, traveled
$\approx 1.9R_\odot$ across the solar disk before crossing the SE
limb; therefore, using a speed estimated by Simnett (2006), it
began at 06:39. This is very close to the eruption in AR~10720,
while no other activity was registered anywhere on the Sun between
06:30 and 07:15. Consequently, this was the opposite side of the
same CME.

Then, assuming the same ratio of the speeds of the NW and SE parts
in the analogous CMEs of 19 and 20 January, and neglecting a small
displacement of the eruption center, one obtains a speed for the
NW component on 20 January of $\approx\,$2075~km\thinspace
s$^{-1}$, or 2585~km\thinspace s$^{-1}$ using the estimate of
Simnett (2006).

The height-time plot for the NW component on 20 January in
Figure~\ref{CME_speeds}b, for a speed of 2075~km\thinspace
s$^{-1}$, has been drawn across the only measured point of
$\approx 4.4 R_\odot$ at 06:54 (LASCO/C2). Our way of estimating
the CME speed is based on the linear fit of sparse data at large
heights, which engenders large uncertainties in the initial parts,
where acceleration occurs. Nevertheless, EIT data points should be
briefly addressed. As noted, they likely correspond to some inner
structures of CMEs; thus, they are expected to be beneath the
height-time plots of the frontal structures. Since the initial
acceleration is large, the initial parts of the height-time plots
for the frontal structures must be curved upward, as Gopalswamy
\textit{et al.} (2005) showed in their Figure~5 (right). The
proximity of EIT points to the linear-fit plots in
Figure~\ref{CME_speeds} for both 19 and 20 January CMEs therefore
supports our estimates of CME speeds.

The NW parts of both CMEs apparently moved faster in the sky plane
and rapidly became faint in their expansion. This is one more
point of similarity between them and explains why the NW component
in the 20 January CME was not detectable in the strongly
contaminated LASCO/C3 images.

Note that shock speeds estimated from drift rates of type II
bursts observed by a few observatories\footnote{ftp:/\negthinspace
/ftp.ngdc.noaa.gov/STP/SOLAR\_DATA/SOLAR\_RADIO/SPECTRAL/
SPEC\_NEW.05} in the analogous events of 15-th (1151\,--\,1900
km\thinspace s$^{-1}$), 17-th (1069\,--\,1578 km\thinspace
s$^{-1}$), 19-th (1093\,--\,1368 km\thinspace s$^{-1}$), and 20-th
(666\,--\,1300 km\thinspace s$^{-1}$) January systematically
decrease, on average, in accordance with the CME speeds measured
from LASCO images (see Table~\ref{Table_720}). Such estimates are
strongly dependent on the coronal density model used and are
therefore highly uncertain. Nevertheless, the shock speeds
estimated for the three preceding CMEs are related to the measured
speeds of the NW components by a factor of 0.4\,--\,0.6 (on
average over observatories). This suggests a systematic factor to
displace the estimated shock speeds with respect to the real ones.
Thus, one might expect the CME speed on 20 January to be even
slightly lower than that of the 19 January CME. In this case, our
evaluation of the CME speed does not appear to be underestimated.

Our estimations are supported by data on the subsequent motion of
the ICME. Measurements from the Solar Mass Ejection Imager
data\footnote{http:/\negthinspace
/creme96.nrl.navy.mil/20Jan05/SMEI.html} indicate a moderately
high ICME speed of $\approx\,$1280 km\thinspace s$^{-1}$ at
elongations of 30\,--\,$60^\circ $. This also corresponds to an
average ICME speed between the Sun and Earth of $\approx\,$1200
km\thinspace s$^{-1}$, if a geomagnetic storm, which started at
about 17 UT on 21 January, was due to its arrival at Earth. By
extrapolating the ``CME vs. ICME speed'' curve shown by Owens and
Cargill (2004, Fig.~3) plotted for the model with aerodynamic drag
(Vr{\v s}nak and Gopalswamy, 2002), one obtains an ICME speed of
1300~km\thinspace s$^{-1}$ corresponding to a CME speed of
2200~km\thinspace s$^{-1}$, while other existing models of ICME
transit between the Sun and Earth provide lesser CME speeds.
Pohjolainen \textit{et al.} (2007) also addressed these issues and
estimated a CME speed of 2500~km\thinspace s$^{-1}$ at
$50R_\odot$, admitting that it may have reached about
3000~km\thinspace s$^{-1}$ in the interval of $(3-50)R_\odot$ and
acknowledging significant uncertainties.

\textit{The CME speed on 20 January was high, but not extreme, and
ranged between 2000 and 2600 km\thinspace s$^{-1}$ in the sky
plane. According to all estimations, at 06:43, when hard emissions
started, the CME frontal structure had reached $>1R_\odot$ above
the solar surface, and the shock was even farther away.}

\section{Relation to SEP and GLE}
\label{SEP_GLE}

\subsection{Comparison with Other Events from the Same Region}

During 15\,--\,19 January, when AR 10720 was already in the
western hemisphere, three other similar major events occurred in
this active region. They included X class flares, strong microwave
bursts, fast CMEs, and significant SEP fluxes. A comparison of all
of these events holds the promise of identifying the origin of the
SEPs. Table~\ref{Table_720} lists their parameters as given in
Solar Geophysical
Data\footnote{http:/\negthinspace/sgd.ngdc.noaa.gov/sgd/jsp/solarindex.jsp}
and the SOHO LASCO CME Catalog.

\begin{table}
\caption{Characteristics of some western major events from AR
10720} \label{Table_720}
\begin{tabular}{cccccccccc} \hline
Date & Time & GOES & Position & \multicolumn{2}{c}{Radio flux, sfu} & \multicolumn{3}{c}{Proton flux, pfu} &  CME \\
  &  &   &  & \multicolumn{2}{c}{$\nu$, GHz} & \multicolumn{3}{c}{$E$, MeV} &  speed \\
  &  &   &  & 8.8  & 15.4  & $>$10 & $>$50 & $>$100 &  km\thinspace s$^{-1}$ \\ \hline

15  & 22:48 &  X2.6 & N14W09 & 20000 & 18000 & 300 & 10 & 0.4 &  2860 \\
17  & 09:43 &  X3.8 & N15W25 & 16000 & 17000 & 3000 & 300 & 25 &  2550 \\
19  & 08:32 &  X1.3 & N15W51 & 17000 & 15000 & $<$100$^a$ & $<$3 & $\approx\,$0.1 &  2020 \\
20  & 06:44 &  X7.1 & N14W61 & 41600 & 77400 & 1800 & 1100 & 680 & 2000\,-- \\
  &   &    &   &   &   &   &   &   & 2600$^b$ \\


\hline
\end{tabular}
$^a$ On top of the background after the preceding event

$^b$ According to the compiled estimations described in the text
\end{table}

The sky-plane CME speeds on 15 and 17 January were even higher
than on 20 January, although AR 10720 was not far from the solar
disk center (thus, their real speeds could well exceed the
projected ones). However, the SEP fluxes from these events were
less than the 20 January event, contrary to the idea that CME
speed is a major factor in determining the SEP productivity of
CME-driven shocks. On the other hand, the SEP fluxes in
Table~\ref{Table_720} roughly correlate with flare emissions, in
particular, at 15.4 GHz. One might ascribe the fact that proton
fluxes were weaker on 15 and 17 January to suboptimal (although
western) position of the flare. However, for the 19 January CME
erupting from W51, the longitude could not appreciably reduce the
proton flux, while its speed was also high. The speeds of all
these CMEs greatly exceeded the Alfv{\'e}n speed at $> 1R_\odot$
above the solar surface (P.~Riley, 2006, private communication),
suggesting that the CME-driven shocks were strong enough to
effectively accelerate particles. Seed populations from previous
CMEs do not appear to be drastically different for 17\,--\,20
January events.

The magnetic connection between the Sun and Earth probably played
a significant role, but proton fluxes and microwaves appear to be
closely related. The strongest fluxes of hard-spectrum protons
near Earth were observed after the 20 January event, when the
flare microwave emission was especially strong and hard. Note also
that the proton flux was rather strong and hard in the 17 January
event, when the microwave peak frequency exceeded 15.4 GHz, as
well as in the 20 January event, unlike all others.

Our comparison also shows that the most distinctive feature of the
20 January event was not a large flux of medium-energy protons,
but an extremely hard proton spectrum (see Table~\ref{Table_720}),
which was a reason for the extreme GLE\,--\,at least, its large
initial part. Flare emissions from the chromosphere and the lower
corona are important indicators of the flare-acceleration of large
quantities of energetic particles. \textit{The magnetic field
strength and the parameters of flare emission appear sufficient to
determine extreme nature of the 20 January 2005 proton event, the
CME speed is not a determining factor.}

\subsection{Gamma-Ray Burst and Ground Level Enhancement}

\begin{figure}    
   \centerline{\includegraphics[width=\textwidth,clip=]{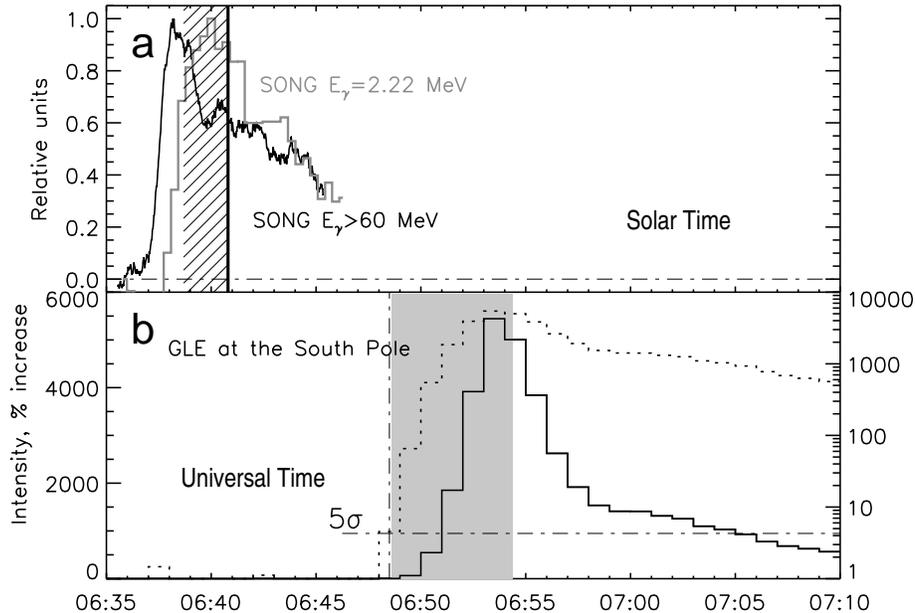}
              }
              \caption{
The impulsive, leading parts of gamma-ray time profiles and GLE.
a)~SONG gamma-ray time profile of $E_\gamma > 60\,$MeV emission
smoothed over 60 points and the 2.22~MeV line emission recorded
with the SONG pulse-height analyzer (gray). The hatched interval
corresponds to the \textbf{latest} possible escape times of
1\,--\,7~GeV protons responsible for the GLE onset. b)~GLE
measured at the South Pole NM station in the linear (solid) and
logarithmic (dotted) scales. Shading marks the arrival interval of
the largest proton flux estimated from the maximum of the
gamma-ray burst.
              }
 \label{GLE_gamma}
   \end{figure}

As noted, the earliest onset and the largest effect of the 20
January 2005 GLE was recorded with the neutron monitor (NM) at the
South Pole (Figure~\ref{GLE_gamma}b). The first protons arrived at
Earth at 06:48:30~UT$\pm$30~s (above $5\sigma$ level, see the
logarithmic scale). They were unusually fast compared to most
GLEs, suggesting a perfect magnetic connection. To determine the
origin of high-energy ($>300$~MeV) protons responsible for the
GLE, we start with a comparison of GLE onset time at the South
Pole with the temporal characteristics of the $\pi^0$-decay
emission (see Kuznetsov \textit{et al.}, 2005a, 2006a,b; Simnett,
2006; Share \textit{et al.}, 2006). As Figure~\ref{X-ray_spectrum}
shows, the $E_\gamma
>60$~MeV gamma-rays are dominated by the $\pi^0$-decay emission.

To estimate the \textit{latest possible escape time} of protons
from/near the Sun, we consider their straight radial flight
between the Sun and Earth by neglecting any delays, curvatures of
their trajectory, and scattering, \textit{i.e.}, conditions
extremely favorable for their prompt transport. The Sun\,--\,Earth
distance was 0.984~AU on 20 January that gives the solar time
scale, $\mathrm{ST} = \mathrm{UT}-489\,\mathrm{s}$\footnote{Solar
Time referring to an event on the Sun and calculated from UT by
taking account of the propagation time}. Plainaki \textit{et al.}
(2007) estimated the effective energy of protons at the GLE onset
to be about 7~GeV, which corresponds to a velocity of $v=0.993c$
and a flight time of 492 seconds. Hence, the protons responsible
for the GLE onset had to depart the Sun \textbf{not later than}
06:40:18~ST$\pm$30~s.

If the energy of the first particles recorded with the South Pole
NM was much less, $\approx 1\,$GeV ($v=0.875c$ and a flight time
559 seconds), then their latest possible escape time was
06:39:11$+$30~s~ST. Thus, allowing for all the uncertainties, the
\textbf{latest} escape time of the first GLE particles was between
06:39:11$-$30~s~ST for 1~GeV and 06:40:18$+$30~s~ST for 7~GeV (a
hatched box in Figure~\ref{GLE_gamma}a). Even this latest-limit
time interval is close to the sharp $\pi^0$-decay burst which
peaked at 06:46\,--\,06:47~UT, \textit{i.e.}, at
06:37:50\,--\,06:38:50~ST. This is the physical limit of the
escape time of the first particles into interplanetary space.

The real path length of charged particles certainly exceeded the
Sun\,--\,Earth distance due to non-zero pitch angles, scattering,
\textit{etc.} Thus, to arrive at Earth by 06:48:30~UT$\pm$30~s,
the first protons had to leave the Sun still earlier than we
estimated. This timing confirms a common origin of the protons
responsible for the high-energy gamma-rays and the leading spike
of the GLE.\footnote{The model injection function of S{\'a}iz
\textit{et al.} (2005) starts later than the physical limit and
therefore should not be used in an analysis of the initial part of
the GLE.}

This conclusion of a common origin for the protons responsible for
the $\pi^0$-decay emission and for the leading GLE spike is
reinforced by comparing their time profiles. These curves resemble
each other; each contains an impulsive, leading part followed by a
long, smoother tail. Their leading parts are much alike in their
rise times, durations, and fast decays; despite broadening of the
GLE due to transport effects. These facts imply that the
processes, which accelerated the particles responsible for the
$\pi^0$-decay emission on one hand, and those responsible for the
leading GLE spike on the other hand, \textit{had proximate
inceptions and comparable increments, acted over almost the same
period, and then shifted from the impulsive mode to another,
prolonged one, in a similar fashion}. The time differences between
these processes did not exceed the uncertainties, and several
reasons exist to account for this difference. An assumption that
these mechanisms were different, independent, distant, and
operated with different populations of particles, seems to be
physically improbable.

The flux of particles at the South Pole reached its maximum while
neutron monitors sensitive to the opposite direction of incident
particles had not yet recorded the GLE. Thus, it had a 100\%
anisotropy, which indicates the direct arrival of particles from
the Sun. The time profile recorded at the South Pole is therefore
well connected with the injection function of solar cosmic rays
influenced by transport effects. On the other hand, the
$\pi^0$-decay emission supplies an approximate representation of
the injection function. The rising phase of the GLE is affected by
dispersion due to differences in path lengths and velocities,
while the fast decay phase is extended due to scattering and the
arrival of particles with large pitch angles. The time profiles of
the $\pi^0$-decay emission and the GLE at the South Pole
(Figure~\ref{GLE_gamma}b) show that transport effects broadened
the GLE, but did not significantly change its shape with respect
to the gamma-ray burst.

If the first 7 GeV GLE protons were accelerated and promptly
escaped during the first $\pi^0$-decay sub-burst (06:43:40~UT or
06:35:31~ST) or the main burst (06:45:30~UT or 06:37:21~ST), then
their real effective path length was between 1.30 and 1.65 of the
Sun\,--\,Earth distance. This effective path length includes all
transport effects, \textit{e.g.}, spiraling due to non-zero
pitch-angles, diffusion time, \textit{etc.} The $\pi^0$-decay
emission maximum (06:38\,--\,06:39~ST) corresponds to the largest
number of high-energy protons in the solar atmosphere. The maximum
of the GLE ($\approx\,$06:53:00) corresponds to the arrival of
most of the 1\,--\,15 GeV protons. Therefore, the arrival times of
protons emitted from the Sun near the $\pi^0$-decay emission
maximum, with our estimated range of path lengths, should
encompass the time of the GLE maximum. For protons of these
energies with our estimated path lengths, arrival times range from
06:48:36 until 06:54:21~UT (shading in Figure~\ref{GLE_gamma}b).
Thus, the $\pi^0$-decay emission and GLE closely correspond to
each other in time, and our estimate of the path length of
high-energy protons is reasonable.

\textit{All these facts, considerations, and estimations
demonstrate that the high-energy particles responsible for
$\pi^0$-decay gamma-rays on the Sun, and the majority of those
responsible for the leading GLE spike, belonged to the same
population, and they acquired a significant part of their energy
from the same acceleration process.} No other realistic
possibility appears to be consistent with all the facts.

\section{Discussion}
\label{Discussion}

\subsection{Overall Picture of the Solar Event}

As shown in Section~\ref{Observations_Analysis}, the great event
of 20 January was primed by the history of the active region,
\textit{i.e.}, the convergence of opposite-polarity domains and
large shear motions. These are known to be precursors of big
flares, which indeed occurred on 15\,--\,20 January. A distinctive
feature of the biggest flare on 20 January was its occurrence just
above sunspot umbrae where the magnetic fields were extremely
strong, $\gsim$3000~G at the photosphere. This also ensured a high
CME speed (cf. {\v S}vestka, 2001). Strong magnetic fields with a
large stored non-potential component were most likely responsible
for the outstanding features of this eruptive flare.

With the flare onset, opposite-polarity magnetic fluxes started to
interact. The magnetic flux in the southern sunspot of N-polarity
was larger. In the course of this interaction (probably via
magnetic reconnection) its lesser counterparts in the northern
sunspot became exhausted, and the process involved counterparts
from the remaining area of the S-polarity domain. This manifested
in an apparent motion of the brightest emission away from the
northern sunspot, whereas the emissions in the southern sunspot
persisted.

Particles accelerated in the corona streamed down to dense layers,
precipitated, and heated them. High-energy electrons produced
gyrosynchrotron emission, which was very strong and hard due to
their large number, their hard spectrum, and strong magnetic
fields. The lost energy was transformed into heating and HXR and
gamma-ray continua as well as H$\alpha$ and 1600~\AA\ emissions.
Heavier particles produced gamma-rays, in particular, in discrete
lines (cf. Ryan 2000; Somov 1992, 2006). The heated chromospheric
plasmas evaporated and filled closed configurations, forming the
hot SXR-emitting loop-like structure and a cooler arcade later on.

RHESSI shows the 2.22 MeV line source to be located on, and
confined to, the northern ribbon where the coronal configuration
was closed. This source was due to the precipitation of
accelerated heavy particles with an energy of at least a few MeV,
that had no direct access into a closed structure from outside.
The hypothetical transport of external particles into the closed
configuration could hardly proceed without other manifestations.
Hence, these heavy particles could be accelerated only within the
magnetic field of the active region, but not outside it. The
absence of a detectable 2.22 MeV line from the southern source
could be due to the suppression of ion precipitation in the
sunspot umbra by mirroring because of much stronger field there
than at the 2.22 MeV line centroid and nearby.

The above items constitute a self-consistent picture of the flare
that encompasses observed phenomena and explains their
particularities. This picture corresponds with many other
observations and well-known concepts.

The CME on 20 January was not distinctive to the other CMEs of
15\,--\,19 January from the same active region, and its speed,
2000\,--\,2600~km\thinspace s$^{-1}$, was not higher than some
others. This is consistent with a conclusion of Reinard and
Andrews (2006) that the only difference between SEP-related and
non-SEP-related CMEs is their 1.36 higher average speed. Both slow
and fast CMEs were present in both groups. On the other hand,
\textit{comparison with other events from AR~10720 demonstrates
that the parameters of the flare alone determine the extreme
nature of the 20 January proton event.}

\subsection{Radio Burst and Proton Event}
\label{RadioDiagnostics}

Both the qualitative and quantitative features of the
electromagnetic emissions of the 20 January 2005 \textit{flare}
are indicative of its proton productivity. West flares with strong
non-impulsive radio bursts at frequencies $\gsim\,$30 GHz are
known to be followed by SEP events; the strongest ones produce
GLEs (Croom, 1970; Cliver, 2006). On the other hand, intense SEP
events usually occur if the microwave burst is strong. Therefore,
the association of the largest GLE with the flare producing the
highest-energy particle flux is not surprising. The huge radio
burst on 20 January exceeds all the criteria for \textit{flare
proton productivity} established in 1970\,--\,1980s, including the
well-known ``U-like spectrum'' (see Croom, 1971; Castelli
\textit{et al}., 1977; Akinyan, Fomichev, and Chertok, 1980a,b,
1981).

\textit{An important test of the flare origin of near-Earth
protons is the comparison of their spectrum, at several tens of
MeV, with the microwave spectrum}. The peak integral spectral
index of protons, roughly estimated from differential GOES-11
channels, is $\approx 1.1-1.3$ in this range, one of the hardest
ever observed. The expected proton index can be estimated from
radio data in three different ways. \textit{i})~From the ratio of
flux densities (Chertok, 1982) observed with NoRP at 9 and 17 GHz,
$F_{9}/F_{17} \approx 0.7$, it is $0.9 F_{9}/F_{17} +0.4 \approx
1$. \textit{ii})~From the spectral maximum of the microwave burst,
30~GHz (Chertok, 1990), which provides a similar proton index.
\textit{iii})~From the parameters of the U-like spectrum (Bakshi
and Barron, 1979) it is $1.185/\log_{10} \left(
\nu_\mathrm{peak}/\nu_{\min} \right) \approx 0.8$. All three
estimates are close to the observed integral spectral index of
protons. With a strong estimated proton flux and very hard
spectrum at several tens of MeV, microwave diagnostics point to a
major proton flux at higher energies, \textit{i.e.}, an intense
GLE.

A glance at lists of radio bursts (\textit{e.g.}, the NoRP Event
List\footnote{http:/\negthinspace/solar.nro.nao.ac.jp/norp/html/event/})
and proton events confirms that strong microwave/millimeter
emissions \textit{e.g.}, with a flux density of $> 10^4$ sfu at
35~GHz, and high turnover frequency are typical of most major
SEP/GLE events, including some extreme proton events of recent
years. However, there are exceptions, \textit{e.g.}, when strong
proton fluxes near Earth were associated with rather weak bursts.
At the same time, the magnetic connectivity between the Sun and
Earth is certainly important. The relationship between strong
microwave bursts and proton events (see also Cliver \textit{et
al.}, 1989) needs further detailed analysis, which is currently in
progress.

Advocating SEP acceleration by coronal shocks, Kahler (1982)
ascribed the correlation between SEP fluxes and the parameters of
microwave bursts to the ``Big Flare Syndrome'', the general
association between a flare's energy release and the magnitude of
its manifestations (also cf. {\v S}vestka, 2001). This tendency is
certainly present, but it could hardly ensure the correspondence
between the spectra of protons near Earth and microwave bursts
(Bakshi and Barron, 1979; Chertok, 1982, 1990). If the role of
coronal shocks in the acceleration of particles is even more
significant, then, as our results demonstrate, energy release in
the source regions of eruptive events, manifesting in flare
emissions, is an important factor in determining the eventual SEP
outcome.

\textit{Powerful microwave/millimeter emissions with high turnover
frequency, radiated by numerous electrons with a hard spectrum in
strong magnetic fields, are typical of major proton flares, but
not of all proton events.}

\subsection{Where Did the Protons Come From?}
\label{where_from}

Both observations and estimations do not obviously rule out the
acceleration of particles in either the flare or the CME-driven
shock. Indeed, if the CME started around 06:36, accelerated in
$\approx 5$ minutes (5\,--\,10~km\thinspace s$^{-2}$) up to a
speed of 2000\,--\,2600~km\thinspace s$^{-1}$ (well in excess of
the Alfv{\'e}n speed), and the piston-driven wave steepened into a
shock in a similar time frame, then the appearance of the shock is
roughly co-temporal with the $\pi^0$-decay burst. At that time,
the CME was at $\approx(2-3)R_\odot$, consistent with
shock-acceleration.

On the other hand, the observations also appear to be consistent
with a flare-related origin of the accelerated protons. As found
in Section~\ref{Observations_Analysis}, sources of electromagnetic
flare emissions produced by accelerated electrons and
moderate-energy protons, from the microwave/millimeter burst up to
the 2.2 MeV line gamma-rays, were compact and located inside a
volume confined by flare ribbons and the SXR-emitting loop-like
structure. The time profiles of these emissions closely matched
each other. Thus, all these emissions were generated within a
compact closed structure above the active region, and the
particles responsible for them were accelerated there also.

Cliver, Kahler, and Vestrand (1993) proposed that particles
accelerated by a CME-driven shock could then precipitate back in
dense layers of the solar atmosphere and produce gamma-rays. In
this scenario, the 2.22 MeV line gamma-ray sources would be
widespread (cf. Hurford \textit{et al.}, 2006a), dominating the
``twin'' dimmings near the ends of a post-eruptive arcade (cf.
Figure~\ref{RHESSI_gamma_images}b), because such dimmings are
considered to be bases of an expanding flux rope, the body of a
CME (Hudson and Webb, 1997; Zarro \textit{et al.}, 1999; Webb
\textit{et al.}, 2000). However, the 2.22 MeV line source was
compact, with its centroid located at a flare ribbon, far from the
dimmings (see Figure~\ref{RHESSI_gamma_images}a). The CME was
located $>1R_{\odot}$ from the active region and so could not
control the detailed course of the flare, nor could the even more
distant shock front.

The observed manifestations of accelerated electrons and protons
match the standard picture of a flare. In this picture, particles
accelerated in the corona stream down along magnetic field lines.
They then precipitate in dense layers and heat them. This causes
emissions from flare ribbons. It also causes chromospheric
evaporation to fill closed coronal structures and produce soft
X-rays (\textit{e.g.}, Somov, 1992; 2006). Our analysis does not
address the escape of particles into interplanetary space, which
could be due to drifts (\textit{e.g.}, in electric fields) of
particles accelerated in the flare region to open magnetic field
lines, which are known to be always present in active regions
(\textit{e.g.}, Fisk and Zurbuchen, 2006). As Cane, Erickson, and
Prestage (2002) showed, open magnetic fields between the low
corona and the Earth do exist, ensuring the transport of electrons
and protons of various energies. Note also the favorable longitude
and proximity to the ecliptic plane ($\approx 9^\circ$) of the
active region on 20 January.

In Section~\ref{SEP_GLE} we established that the high-energy
particles responsible for the $\pi^0$-decay gamma-rays and the
majority of those responsible for the leading spike of the GLE
belonged to the same population and were accelerated by the same
mechanism. As shown in Section~\ref{HXR_gamma_timeprofiles}, the
time profiles of the $\pi^0$-decay emission and lower-energy
emissions (in particular, electron bremsstrahlung) were close in
time and roughly similar, but still different. These
dissimilarities might mean that these emissions were generated by
either (i)~two different particle components (G.H.~Share
\textit{et al.}, 2007, private communication) or (ii)~the same
population with variable parameters. In the former case, a reason
could be (a)~an extra process, which started to operate in the
flare region slightly later and accelerated heavy particles up to
$> 300\,$MeV, and (b)~another source located at a distance,
including a CME-driven shock.

However, the single-population case is favored by the
correspondence of the structural details in time profiles of the
$\pi^0$-decay and lower-energy emissions including bremsstrahlung
(Figure~\ref{RHESSI_SONG_timeprofs}) and microwaves
(Figure~\ref{norp_timeprofs}d,e; \textit{e.g.}, peaks at 06:46 and
06:47:15), with their relative magnitudes corresponding to a
continuous spectrum. The relative timing of the $\pi^0$-decay and
lower-energy bursts is consistent with variations in the energy
release rate in the flare. The energy release rate, in turn, is
directly related to the energy and quantity of accelerated
particles. Hard electromagnetic emissions (in particular, the
$\pi^0$-decay one) peak around 06:46, as did the microwave burst.
At that time, the flare was located in the strongest magnetic
fields (Sections~\ref{TRACE_RHESSI_analysis}, \ref{MW_burst}), and
the energy, and number, of accelerated particles reached their
maxima. Thus, the peak time of the $\pi^0$-decay burst is
consistent with a flare origin, where it would be expected to
start later than the lower-energy emissions, \textit{i.e.}, when
the energy release rate became high enough.

Correspondence between the parameters of the (flare originating)
microwave/ millimeter burst with those of the protons responsible
for the initial, impulsive part of the SEP/GLE, including such a
delicate characteristic as a hard energy spectrum of protons
(Section~\ref{RadioDiagnostics}), supports flare-related
acceleration for this part of the SEP/GLE. Only the flare, not the
CME, had extreme characteristics comparable with those of the 20
January 2005 proton event. Note that from a similar analysis of a
few other events, Li \textit{et al.} (2007a,b) also concluded that
the particles responsible for the impulsive, initial component of
GLEs were accelerated in flare current sheets.

Our results favor acceleration of protons just in the flare. On
the other hand, the available data do not permit us to determine
if the GLE-productive particles acquired energy up to $\gsim
7\,$GeV in the flare, or by an additional acceleration mechanism,
\textit{e.g.}, by a CME-driven shock (see Desai \textit{et al.},
2006). A posterior extra shock-acceleration is favored if the
particles constituting a seed population (see Tylka \textit{et
al.}, 2005) collide with its convex front from outside with a
small angle. However, this mechanism does not appear to be
efficient, if particles collide with it from inside, which was
most likely the case during the main flare phase of the 20 January
2005 event.

\subsection{Comments on Later Manifestations and SEP Properties}

Our considerations addressed the main stage of the flare and the
related initial GLE spike to minimize secondary factors and
transport effects in interplanetary space, which might influence
SEP properties (\textit{e.g.}, Cane \textit{et al.}, 2003, 2006;
Ng, Reames, and Tylka, 2003; Tylka \textit{et al.}, 2005). Here we
note that, in its next orbit, RHESSI registered gamma-rays
identified with the $\pi^0$-decay emission with a decay time of
$\tau_{\pi^0} \approx 100$ min, while the 2.22 MeV line emission
(see Figure~\ref{GLE_gamma}a) and nuclear de-excitation emissions
decayed with $\tau_{2.22\ \mathrm{MeV}} \lsim 10$ min (Share
\textit{et al.}, 2006; G.H.~Share, 2007, private communication).
It is unclear if long-decay gamma-ray emissions are generally due
to prolonged acceleration or storage of impulsively accelerated
protons (Ryan, 2000). In our event, a CME-driven shock cannot be
responsible for gamma-rays with $\tau_{\pi^0} \approx 100$ min.
Such an acceleration in the active region should accelerate
electrons as well, but Learmonth
records\footnote{http:/\negthinspace
/www.ngdc.noaa.gov/stp/SOLAR/ftpsolarradio.html} show no evidence
of this in the radio domain. If the long gamma-ray emissions were
due to trapped protons then the decay times would be limited by
their collision rate. The mean free path $\lambda_{\mathrm i}$ of
a fast ion with a mass $m_{\mathrm i}$, charge $e_{\mathrm i}$,
and an initial velocity $v_0$, in a plasma with a number density
$n$, is $\lambda_{\mathrm i} = m_{\mathrm i} m_{\mathrm e}
v{_0}{^4}/ \left( 16 \pi e_{\mathrm i} e_{\mathrm e} \Lambda
n\right)$, where $\Lambda \approx 10$ Coulomb logarithm,
$e_{\mathrm e}$ and $m_{\mathrm e}$ are the electron charge and
mass; $\tau_{\mathrm{coll}} \approx \lambda_{\mathrm i}/v_0$. With
a density in the SXR-emitting loop-like structure of $\approx
10^{11}\,$cm$^{-3}$, $\tau \approx 80$ minutes for 300 MeV protons
and $\tau \approx 4$ minutes for 30 MeV protons are close to the
observed parameters. On the other hand, with $n < 10^9\,$cm$^{-3}$
expected in the corona at heights $> 1 R_{\odot}$, the collision
time for 300 MeV protons is very long, $\tau > 120$ hours. Thus,
the observed decay times of gamma-ray emissions appear to be
consistent with the trapping of protons in the observed flare
loop.

The major conclusion drawn from analyses of the SEP composition,
based on observations during this SEP
event\footnote{http:/\negthinspace/creme96.nrl.navy.mil/20Jan05/},
is its correspondence to ``gradual events'', which are believed to
be due to acceleration in CME-driven shocks (Reames, 1995), rather
than in a flare. However, Labrador \textit{et al.} (2005) inferred
an acceleration time of $< 90$~s, which appears to be consistent
with the time profiles of gamma-ray emissions. R.A.~Mewaldt (2007,
private communication) found that the measured variations of the
Fe/O ratio might be indicative of an impulsive flare-related
component along with a longer gradual one, as suggested by Cane
\textit{et al.} (2003, 2006). However, as Ng, Reames, and Tylka
(2003) showed, initial enhancements of the Fe/O ratio can be due
to transport effects. Distinguishing all these effects does not
seem to be a simple task.

Thus, the situation with the SEP composition is not obvious.
Analyses of the corresponding data (especially on such short
timescales as the leading SEP/GLE spike), as well as related
issues, probably require the consideration of several complex
factors. A simplified glance at these data might lead to mistaken
conclusions about the origin of SEPs. Furthermore, as Somov and
Chertok (1996)\footnote{Chapman Conference ``Coronal mass
ejections: causes and consequences'', Montana State University,
Bozeman, Montana} pointed out, the observed properties of SEPs
might be dependent on plasma parameters at the acceleration site.
In particular, if the acceleration of ions occurs in a low-density
region above the flare site, relatively high in the corona, then
the number of Coulomb collisions might be too small to ensure
their Maxwellian distribution, which is inherent in the
temperature of the acceleration site. As a result, these SEPs
could retain properties typical of background coronal conditions.

To determine if the probable contribution from the CME-driven
shock to the acceleration of heavier ions was as significant as
the contribution from the flare-accelerated protons to the leading
spike of the GLE, a comparison should be made with other events,
especially with those from the same active region. In this
comparison, not only the SEP composition, but also total fluxes
are important. Otherwise, there is no apparent reason to relate
the outstanding properties of the 20 January 2005 event with the
parameters of the CME only.

\section{Summary and Conclusion}
\label{Summary}

In considering the extreme proton event of 20 January 2005, we
first carried out a comprehensive analysis of the flare, based on
multi-spectral observations, and second, investigated the origin
of the energetic protons corresponding to the leading SEP/GLE
spike. Our conclusions regarding the flare are as follows.

\begin{enumerate}

\item Imaging data in H$\alpha$, 1600~\AA, soft and hard X-rays,
EUV, and 2.22 MeV neutron-capture line show that the sources of
these emissions were compact and localized within a volume
confined by flare ribbons, which crossed the umbrae of the largest
sunspots with strong magnetic fields.

\item Non-imaging data extend this conclusion to
microwave/millimeter, lower-energy, and medium-energy gamma-ray
emissions, because the time profiles of the major flare component,
caused by accelerated electrons and protons, were similar and
closely related, particularly in time.

\item The $\pi^0$-decay emission time profile was much like those
of lower energy, but their differences prevent us from confidently
asserting a common source. Nevertheless, co-temporal structural
details in the $\pi^0$-decay and lower-energy time profiles (their
magnitude corresponds to a continuous spectrum) indicate their
common nature; the timing of the $\pi^0$-decay gamma-ray burst
corresponds to the highest energy release rate in the flare
(Section~\ref{where_from}). Thus, its flare origin appears to be
preferable. However, one should not consider the flare origin to
be proven.

\item The above considerations favor the acceleration of both the
electrons and protons, responsible for various emissions (at
least, at moderate energies), within the magnetic field of the
active region at a moderate height above the sunspots. The
manifestations of the electron and proton components correspond to
the standard picture of a flare, in which chromospheric ribbons
and compact sources of hard emissions map the footpoints of newly
formed magnetic loops, along which energetic particles fly
downward and bombard the lower atmosphere.

\item The presence of a large number of accelerated particles in
strong magnetic fields is demonstrated by the fact that the flare
was accompanied by one of the strongest and hardest-spectrum
microwave bursts ever observed, with a peak flux density of almost
$10^{5}$ sfu at 30 GHz and a very strong ($5\times 10^{4}$ sfu)
emission at 80 GHz.

\item Evaluations of the 20 January CME sky-plane speed from
various available data result in an estimate of 2000\,--\,2600
km\thinspace s$^{-1}$.

\end{enumerate}

Regarding the SEP/GLE aspect of this event, we first note that
there is no clear observational evidence to state with certainty
if it was due to the flare or the CME-driven shock. However, our
analysis favors the flare region as the probable site of
acceleration of the particles responsible for the leading SEP/GLE
spike, rather than the shock ahead of the CME, as supported by the
following arguments.

\begin{enumerate}

\item The compactness of the gamma-ray sources and similarity of
the time profiles of different emissions could hardly be expected
if high-energy particles were accelerated in an extended shock
front ahead of the receding CME.

\item The proton event under consideration was extreme, not in its
CME speed, but in the parameters of the flare itself, particularly
in the huge microwave/millim\-eter radio burst.

\item There is a significant correspondence between the hardness
of the microwave spectrum and the hardness of the energetic
spectrum of the SEP and GLE protons.

\item The time of emission from the Sun of high-energy protons
responsible for the onset of the GLE is close to that of the
$\pi^0$-decay emission.

\item The gamma-ray burst and the leading GLE spike have
similar shapes.

\item The decay times of gamma-ray emissions computed by
G.H.~Share (2007, private communication) from RHESSI data agree
with the collision times of protons trapped in the observed flare
loop, but not with those stored higher in the corona.

\end{enumerate}

The available data, together with our results, do not rule out a
possible role for the CME-driven shock in the acceleration of
particles during the initial part of the GLE. Similarly, the role
of flare acceleration cannot be completely ruled out for later
stages of the SEP event, when the shock seems to dominate.
Advocating ``shock-acceleration only'' or ``flare-acceleration
only'' does not seem to be productive (see {\v S}vestka, 2001;
Kallenrode, 2003). Studies of the relative contribution from
acceleration in flares and CME-driven shocks appears to be more
fruitful. They can vary for dissimilar events, their different
stages, and in different energy ranges.

The 20 January 2005 event represents a distinct class of extreme
solar flares. They amount to a small percentage of all flares, but
probably constitute the majority of events which can be
proton-rich under favorable Sun\,--\,Earth connections. The
expected features of these flares are: occurrence near or above
sunspot umbrae in strong magnetic fields; powerful bursts in
microwaves; and at long millimeter wavelengths. These events are
very dangerous due to a high probability of strong proton fluxes
with hard spectra. This fact highlights the importance of
measuring strong magnetic fields in solar active regions, as well
as patrol observations of the total radio flux at long millimeter
wavelengths, for the forecast and diagnosis of major proton
events. Currently, only NoRP observations at 35 and 80~GHz are
available from $\approx\,$22 till $\approx\,$08~UT.

\acknowledgements

We appreciate discussions and assistance of A.J. Tylka, G.H.
Share, G.J. Hurford, M.A. Livshits, G.V. Rudenko, I.I. Myshyakov,
S.A. Bogachev, A.B. Struminsky, S.M. White, B.R. Dennis, and A.K.
Tolbert. We are grateful to G.H. Share and G.J. Hurford for
supplying us with preliminary RHESSI data. We thank the anonymous
reviewers for useful remarks.

The CME catalog is generated and maintained at the CDAW Data
Center by NASA and The Catholic University of America in
cooperation with the Naval Research Laboratory. We thank the teams
of EIT, LASCO, MDI on SOHO (ESA \& NASA); the USAF RSTN Network,
and the GOES satellites.

The study is supported by the Russian Foundation of Basic Research
(05-02-17487, 06-02-16106, 06-02-16239, 06-02-16295, and
07-02-00101), the Federal Ministry of Education and Science
(8499.2006.2, 4573.2008.2, and UR.02.02.509/ 05-1), and the
programs of the Russian Academy of Sciences ``Solar Activity and
Physical Processes in the Sun-Earth System'' and ``Plasma
Heliophysics''. VG is indebted to the SHINE and the US National
Science Foundation for providing support to attend the 2007
Workshop. Discussions held there greatly helped in illuminating
different aspects of problems in question.

\end{article}

\end{document}